\documentclass[numberedappendix,letter]{emulateapj}
\newcommand{\LCDM}{$\Lambda$CDM}
\newcommand{\be}{\begin{equation}}
\newcommand{\ee}{\end{equation}}
\newcommand{\ba}{\begin{eqnarray}}
\newcommand{\ea}{\end{eqnarray}}
\def\ltsima{$\; \buildrel < \over \sim \;$}
\def\ltsim{\lower.5ex\hbox{\ltsima}}
\def\gtsima{$\; \buildrel > \over \sim \;$}
\def\gtsim{\lower.5ex\hbox{\gtsima}}

\newcommand{\wmap}    {{\sl WMAP}}
\newcommand{\planck}    {{\sl Planck}}
\newcommand{\mat}[1]{\ensuremath{\mathbf #1}}   
\newcommand{\arone}{148\,GHz}
\newcommand{\artwo}{218\,GHz}
\newcommand{\uk}  
           {\ensuremath{\mu K}}
\usepackage{graphicx}        
\usepackage{bm}              
\usepackage{natbib}
\usepackage{mathrsfs}

\shorttitle{ACT Likelihood}
\shortauthors{Dunkley, Calabrese, Sievers et al.}

\begin{document}

\title{The Atacama Cosmology Telescope: likelihood for small-scale CMB data}

\begin{abstract}
The Atacama Cosmology Telescope has measured the angular power spectra of microwave fluctuations to arcminute scales at frequencies of 148 and 218 GHz, from three seasons of data. At small scales the fluctuations in the primordial Cosmic Microwave Background (CMB) become increasingly obscured by extragalactic foregounds and secondary CMB signals. We present results from a nine-parameter model describing these secondary effects, including the thermal and kinematic Sunyaev-Zel'dovich (tSZ and kSZ) power; the clustered and Poisson-like power from Cosmic Infrared Background (CIB) sources, and their frequency scaling; the tSZ-CIB correlation coefficient; the extragalactic radio source power; and thermal dust emission from Galactic cirrus in two different regions of the sky. In order to extract cosmological parameters, we describe a likelihood function for the ACT data, fitting this model to the multi-frequency spectra in the multipole range $500<\ell<10000$. We extend the likelihood to include spectra from the South Pole Telescope at frequencies of 95, 150, and 220~GHz. Accounting for different radio source levels and Galactic cirrus emission, the same model provides an excellent fit to both datasets simultaneously, with $\chi^2$/dof= $675/697$ for ACT, and $96/107$ for SPT. We then use the multi-frequency likelihood to estimate the CMB power spectrum from ACT in bandpowers, marginalizing over the secondary parameters. This provides a simplified `CMB-only' likelihood in the range $500<\ell<3500$ for use in cosmological parameter estimation.
\end{abstract}


\author{
J.~Dunkley\altaffilmark{1},
E.~Calabrese\altaffilmark{1},
J.~Sievers\altaffilmark{2},
G.~E.~Addison\altaffilmark{3,1},
N.~Battaglia\altaffilmark{4},
E.~S.~Battistelli\altaffilmark{5},
J.~R.~Bond\altaffilmark{6},
S.~Das\altaffilmark{7,8},
M.~J.~Devlin\altaffilmark{9},
R.~D\"{u}nner\altaffilmark{10},
J.~W.~Fowler\altaffilmark{11},
M.~Gralla\altaffilmark{12},
A.~Hajian\altaffilmark{6},
M.~Halpern\altaffilmark{3},
M.~Hasselfield\altaffilmark{13,3},
A.~D.~Hincks\altaffilmark{6},
R.~Hlozek\altaffilmark{13,2},
J.~P.~Hughes\altaffilmark{14},
K.~D.~Irwin\altaffilmark{11},
A.~Kosowsky\altaffilmark{15},
T.~Louis\altaffilmark{1},
T.~A.~Marriage\altaffilmark{12,13,2},
D.~Marsden\altaffilmark{9,16},
F.~Menanteau\altaffilmark{14},
K.~Moodley\altaffilmark{17},
M.~Niemack\altaffilmark{18,11}
M.~R.~Nolta\altaffilmark{6},
L.~A.~Page\altaffilmark{2},
B.~Partridge\altaffilmark{19}
N.~Sehgal\altaffilmark{20},
D.~N.~Spergel\altaffilmark{13},
S.~T.~Staggs\altaffilmark{2},
E.~R.~Switzer\altaffilmark{6},
H.~Trac\altaffilmark{4},
E.~Wollack\altaffilmark{21}}

\affiliation{$^{1}$ Sub-department of Astrophysics, University of Oxford, Keble Road, Oxford OX1 3RH, UK}
\affiliation{$^{2}$ Joseph Henry Laboratories of Physics, Jadwin Hall,
Princeton University, Princeton, NJ, USA 08544}
\affiliation{$^{3}$ Department of Physics and Astronomy, University of
British Columbia, Vancouver, BC, Canada V6T 1Z4}
\affiliation{$^{4}$ McWilliams Center for Cosmology, Wean Hall, Carnegie Mellon University, 5000 Forbes Ave., Pittsburgh PA 15213, USA}
\affiliation{$^{5}$ Department of Physics, University of Rome `La Sapienza', Piazzale Aldo Moro 5, I-00185 Rome, Italy}
\affiliation{$^{6}$ Canadian Institute for Theoretical Astrophysics, University of
Toronto, Toronto, ON, Canada M5S 3H8}
\affiliation{$^{7}$ High Energy Physics Division, Argonne National Laboratory, 
9700 S Cass Avenue, Lemont IL 60439, USA}
\affiliation{$^{8}$ Berkeley Center for Cosmological Physics, LBL and
Department of Physics, University of California, Berkeley, CA, USA 94720}
\affiliation{$^{9}$ Department of Physics and Astronomy, University of
Pennsylvania, 209 South 33rd Street, Philadelphia, PA, USA 19104}
\affiliation{$^{10}$ Departamento de Astronom{\'{i}}a y Astrof{\'{i}}sica, Pontific\'{i}a Universidad Cat\'{o}lica de Chile, Casilla 306, Santiago 22, Chile}
\affiliation{$^{11}$ NIST Quantum Devices Group, 325
Broadway Mailcode 817.03, Boulder, CO, USA 80305}
\affiliation{$^{12}$ Dept. of Physics and Astronomy, The Johns Hopkins University, 3400 N. Charles St., Baltimore, MD 21218-2686, USA}
\affiliation{$^{13}$ Department of Astrophysical Sciences, Peyton Hall, 
Princeton University, Princeton, NJ USA 08544}
\affiliation{$^{14}$ Department of Physics and Astronomy, Rutgers, 
The State University of New Jersey, Piscataway, NJ USA 08854-8019}
\affiliation{$^{15}$ Department of Physics and Astronomy, University of Pittsburgh, Pittsburgh, PA, USA 15260}
\affiliation{$^{16}$ Department of Physics, University of California Santa Barbara, CA 93106, USA}
\affiliation{$^{17}$ Astrophysics and Cosmology Research Unit, School of
Mathematical Sciences, University of KwaZulu-Natal, Durban, 4041,
South Africa}
\affiliation{$^{18}$ Department of Physics, Cornell University, Ithaca, NY, USA 14853}
\affiliation{$^{19}$ Department of Physics and Astronomy, Haverford College,
Haverford, PA, USA 19041}
\affiliation{$^{20}$ Stony Brook University, Physics and Astronomy Department, Stony Brook, NY, USA 11794}
\affiliation{$^{21}$ NASA/Goddard Space Flight Center, Greenbelt, MD, USA 20771}
\maketitle

\section{Introduction}
\label{sec:intro}
Measurements of the Cosmic Microwave Background (CMB) have played a central role in constraining cosmological models. Anisotropies measured over the whole sky by \wmap\ have provided evidence for a flat universe described by just six cosmological parameters. The measurement of the Sachs-Wolfe plateau in the power spectrum, and three acoustic peaks, have led to constraints on \LCDM\ parameters to percent-level accuracy \citep{komatsu/etal:2011,larson/etal:2011}. The Silk damping tail of the power spectrum provides a wealth of additional information about the physics of the early universe, encoded in its shape, and in the positions and heights of the higher-order acoustic peaks \citep{silk:1968}. Extracting information from these angular scales is complicated by the presence of additional power from extragalactic point sources, emission from the Galaxy, and secondary anisotropies due to the thermal and kinematic Sunyaev-Zel'dovich effects \citep{sunyaev/zeldovich:1970}.

The Atacama Cosmology Telescope (ACT) mapped the mm-wave sky with arcminute resolution from 2007 to 2010 in two distinct areas. About 600 square degrees were used to compute the angular power spectrum. Power spectra and cosmological results using the 1-year data, from the 2008 observing season, were presented in \citet{fowler/etal:2010,das/etal:2011}; and \citet{dunkley/etal:2011}. During roughly the same period, the South Pole Telescope also mapped the microwave sky, and presented cosmological results in \citet{lueker/etal:2010,shirokoff/etal:2011,keisler/etal:2011,reichardt/etal:2012}; and \citet{story/etal:prep}.

In this paper we describe a method to fit multi-frequency power spectra from the ACT data simultaneously for CMB, foreground, and SZ parameters, following a similar approach to analyses in \citet{dunkley/etal:2011} and \citet{reichardt/etal:2012}. We describe the likelihood constructed for the 3-year ACT dataset, using data from the 2008-2010 observing seasons, and show how it can be used in combination with data from SPT in a self-consistent way. Using this likelihood from ACT, we then construct a simpler CMB-only likelihood, estimating CMB bandpowers marginalized over the SZ and foreground parameters.

This is one of a set of papers on the 3-year ACT data; \citet{das/etal:prep} present the angular power spectra, and \citet{sievers/etal:prep} use the likelihoods presented here to estimate cosmological parameters. We begin in \S \ref{sec:methods} by describing the model for the mm-wave emission. In \S \ref{sec:like} the full likelihood of the ACT data is described, including the combination with data from SPT and \wmap\ 7-year data. In \S \ref{sec:results} we show the small-scale model fit to the multi-frequency data. In \S \ref{sec:cmb} we describe the compressed CMB-only likelihood, concluding in \S\ref{sec:conclude}.

\section{Model for the mm-wave sky}
\label{sec:methods}

\begin{figure*}
\centering
\epsscale{0.9} 
\plotone{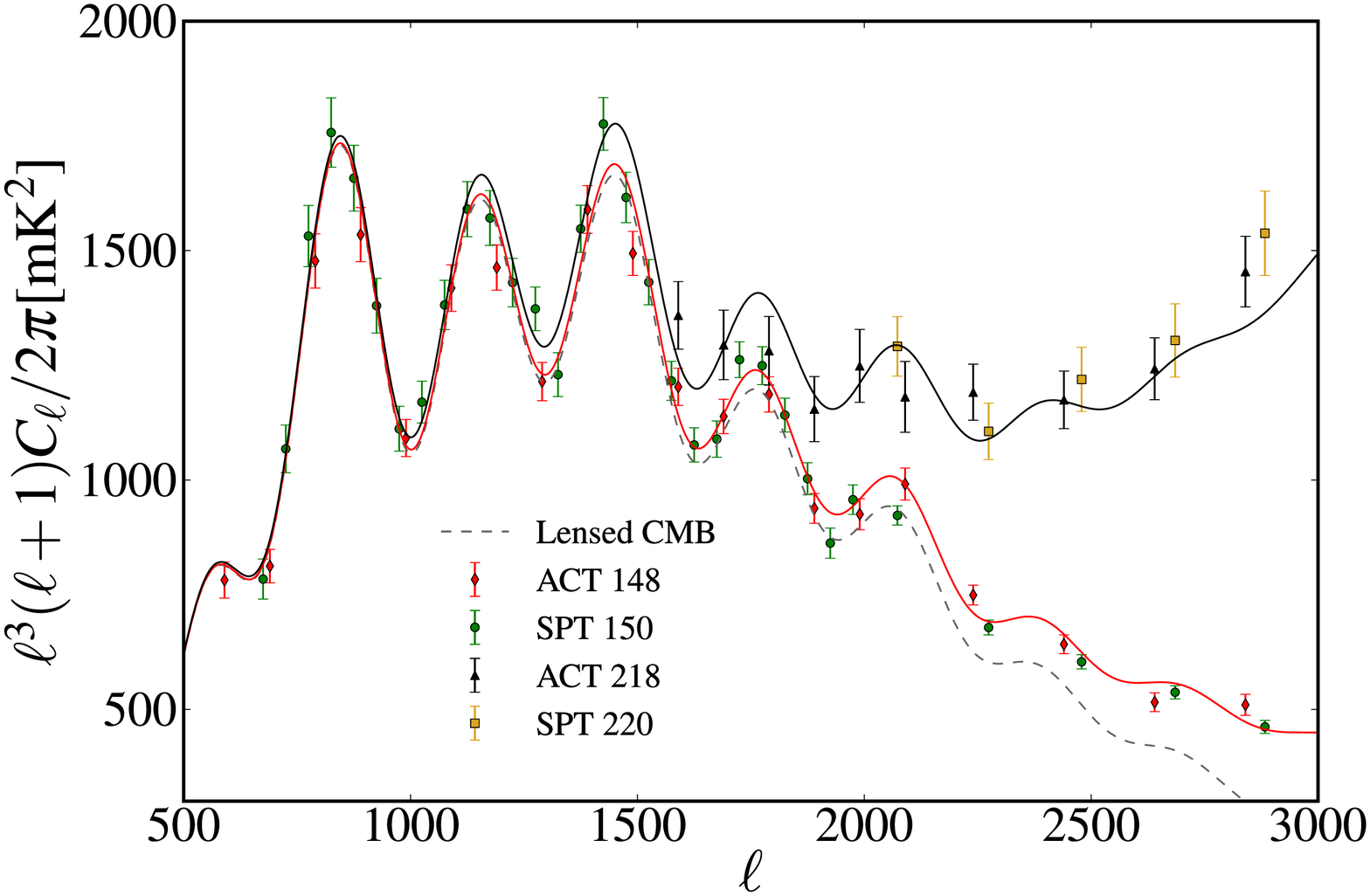}
\caption{Summary of small-scale mm-wave data measured by the Atacama Cosmology Telescope \citep{das/etal:prep} and the South Pole Telescope \citep{keisler/etal:2011,reichardt/etal:2012}, in the angular range used for measuring the damping tail of the CMB. The ACT and SPT data are independently calibrated to \wmap. The vertical axis is $\ell^4 C_\ell$ instead of the conventional $\ell^2 C_\ell$ to highlight the features at these angular scales. The primary CMB signal corresponding to the best-fitting \LCDM\ model \citep{sievers/etal:prep} is indicated (dashed), together with the total signal at 148~GHz (red, lower solid curve) and 217~GHz (black, upper solid curve), including secondary effects from SZ and foregrounds. Modeling the secondary contributions from SZ and foregrounds is vital to allow extraction of the primordial signal at small scales. 
\label{fig:l4}}
\end{figure*}

\begin{table*} [t]
\centering
\caption{Small-scale CMB datasets}
\begin{tabular}{ccccccccccc}
\hline
Dataset & Frequency\tablenotemark{a} & Reference & Area & $\ell_{\rm min}$ & $\ell_{\rm max}$  & $S_c$ \tablenotemark{b} &  $\nu_{\rm tSZ}$\tablenotemark{c} & $\nu_{\rm Rad}$ & $\nu_{\rm CIB}$ \\
& GHz &  & sq degrees &  &   & mJy &  GHz & GHz & GHz\\
\hline
ACT  & 148 & \citet{das/etal:prep} & 590\tablenotemark{d}  & 500 & 10000  & 15  & 146.9 & 147.6 & 149.7 \\
                                    & 218 &  & & 1500 & 10000 &   & 220.2 & 217.6 & 219.6 \\
\hline
SPT-low  & 150 & \citet{keisler/etal:2011} & 790 & 650 & 2000 & 50 &152.9 & 150.2 & 153.8 \\
\\
SPT-high  & 95 & \citet{reichardt/etal:2012} & 800 & 2000 & 9400 &   & 97.6 & 95.3 & 97.9 \\
                                         & 150 &  & & 2000 & 9400 & 6.4  & 152.9 & 150.2 & 153.8 \\
                                        & 220 &  & & 2000 & 9400  &  & 218.1 & 214.1 & 219.6 \\
\hline
\footnotetext[1]{All cross-spectra between channels are used in the likelihood.} 
\footnotetext[2]{Flux cut imposed on map by its point source mask.}
\footnotetext[3]{Effective band-centers from ACT are from \citet{swetz/etal:2011}, given for tSZ, radio sources, and CIB sources.}
\footnotetext[4]{This area includes the ACT-E region at dec~$=0^\circ$ (300 deg$^2$), and the ACT-S region at dec~$=-55^\circ$ (290 deg$^2$).}
\end{tabular}
\label{table:data}
\end{table*}

\begin{figure}
\vskip -0.5cm
\epsscale{1} 
\centering
\includegraphics[scale=.33,angle=90]{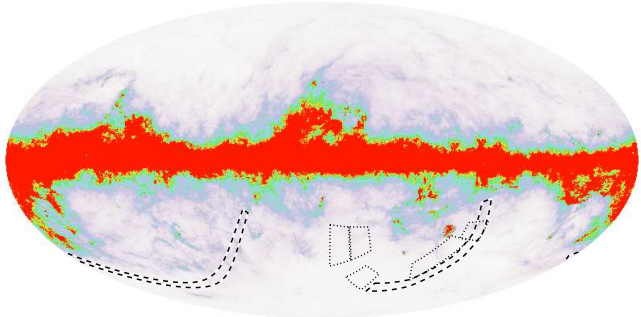}
\caption{ Regions of the sky used for ACT power spectra \citep[dashed,][]{das/etal:prep}  in the Equatorial plane (ACT-E, 300 deg$^2$), and at -55$^\circ$ declination (ACT-S, 292 deg$^2$). The 800 deg$^2$ used for SPT power spectra \citep[dotted,][]{keisler/etal:2011,reichardt/etal:2012} is indicated, with 54 deg$^2$ overlap with ACT-S. The color scales with Galactic cirrus intensity \citep{finkbeiner/davis/schlegel:1999}.
\label{fig:sky_area}
}
\end{figure}

Sky maps of mm-wave fluctuations at arcminute resolution include components emitting at low redshift, in addition to the primordial CMB signal and secondary CMB effects \citep[e.g.,][]{sievers/etal:2009,fowler/etal:2010,dunkley/etal:2011,keisler/etal:2011}. The power spectra from the complete ACT dataset, reported in \citet{das/etal:prep}, are shown in Figure \ref{fig:l4}, focusing on angular scales of interest for the primordial CMB signal. At scales smaller than a few arcminutes ($\ell\simeq 1500$) the secondary signal, which we define as the sum of foregrounds and SZ effects, becomes significant compared to the CMB. We want to extract the primary CMB signal, but, since there are more foreground components than frequency channels, information about both the frequency and scale dependence of the foregrounds is required to separate the signals. In this section we describe a model to fit the power spectrum of these fluctuations over the frequency range $90\ltsim \nu \ltsim 250$~GHz probed by ACT, SPT, and other CMB experiments including the \planck\ satellite \citep{tauber/etal:2011}. We follow a similar approach to \citet{sievers/etal:2009,dunkley/etal:2011,reichardt/etal:2012}. 

For frequency $\nu$ and direction ${\bf\hat{n}}$ we model the signal in the maps as
\be
\Delta T(\nu,{\bf\hat{n}}) = \Delta T^{\rm CMB}({\bf\hat{n}}) + \Delta T^{\rm sec}(\nu,{\bf\hat{n}}),
\label{eqn:map_model}
\ee
where $\Delta T^{\rm CMB}({\bf\hat{n}})$ are the lensed CMB fluctuations, which are independent of frequency in thermodynamic units. The secondary signal, $\Delta T^{\rm sec}(\nu,{\bf\hat{n}})$, is dominated by the sum of tSZ and kSZ components, emission from dusty infrared galaxies and radio galaxies, and dust emission from Galactic cirrus, all of which are functions of frequency.

The cross-correlation power spectra between frequency $\nu_i$ and $\nu_j$ are 
calculated as
\be
C^{\rm ij}_\ell=\left<{\tilde T}^*_\ell(\nu_{\rm i}){\tilde T}_\ell(\nu_{\rm j})\right>,
\ee
where ${\tilde T}_\ell$ is the Fourier transform of $T({\bf\hat n})$ in the flat-sky approximation. 
The theoretical cross-spectrum ${\cal B}_\ell^{\rm{th,ij}}\equiv \ell(\ell+1){C}_\ell^{\rm{th,ij}}/2\pi$ is modeled as 
\be
{\cal B}_\ell^{\rm{th,ij}} = {\cal B}_\ell^{\rm{CMB}} + {\cal B}_\ell^{\rm{sec,ij}}, 
\ee
where  ${\cal B}_\ell^{\rm {CMB}}$ is the lensed primary CMB power spectrum. In this analysis we model the secondary spectra as
\ba
\label{eq:model}
{\cal B}_\ell^{\rm{sec,ij}}  &=&  {\cal B}_\ell^{\rm{tSZ,ij}} 
+ {\cal B}_\ell^{\rm{kSZ,ij}}+ {\cal B}_\ell^{\rm{CIB-P,ij}} + {\cal B}_\ell^{\rm{CIB-C,ij}} \nonumber \\
&&
+ {\cal B}_\ell^{\rm{tSZ-CIB,ij}}
+  {\cal B}_\ell^{\rm{rad,ij}}
+ {\cal B}_\ell^{\rm{Gal,ij}},
\label{eqn:spectra_th}
\ea
with contributions from the tSZ and kSZ effects; dusty galaxies that form part of the Cosmic Infrared Background, both Poisson-like (CIB-P) and clustered (CIB-C); the cross-correlation between the tSZ effect and the CIB (tSZ-CIB); radio galaxies (rad); and dust emission from Galactic cirrus (Gal). We assume that all other cross-spectra can be neglected. Measurements by \wmap\ and other CMB experiments show that other Galactic emission, including synchrotron and free-free emission, is negligible in the $\nu \gtsim 90$~GHz frequency range at these small scales and locations \citep[e.g.,][]{gold/etal:2011}. The cross-correlation of radio sources and both tSZ and CIB sources is also expected to be small \citep[e.g.,][who find a correlation of only a few per cent in simulations]{sehgal/etal:2010a}.
Since the kSZ signal consists of positive and negative fluctuations, depending on the line-of-sight motion of the electrons that source the signal, the two-point correlation function with other signals should average to zero. 

The majority of these secondary spectra will be common to all regions of the sky and to different experiments. The power in the residual radio point sources is expected, however, to vary among data sets due to the removal of bright sources. For example, for a Poisson distribution of sources with differential number counts scaling as $dN/dS \propto S^{-2}$, the Poisson power will be 
\be
C_\ell=\int_0^{S_{\rm max}}S^2\frac{dN}{dS}dS,
\ee
i.e., $C_\ell \propto S_{\rm max}$, where $S_{\rm max}$ is the flux of the brightest sources in the map at a given frequency. Deeper surveys, with lower noise per pixel, are able to detect and mask out dimmer sources, so $S_{\rm max}$ will be lower, leading to a lower residual power. Radio sources have a shallower $dN/dS$ slope than CIB galaxies, so imposing a flux cut of, e.g., $S_{\rm max}= 15~$mJy removes a significant amount of radio power, but little CIB power. 

The Galactic foreground power is also expected to vary between regions on the sky. The two regions mapped by ACT (Equatorial, ACT-E, and South, ACT-S) are shown in Figure \ref{fig:sky_area} and summarized in Table \ref{table:data}, together with data from SPT. The temperature scale of the Galactic cirrus map  from \citet{finkbeiner/davis/schlegel:1999} is shown for comparison. A higher level of emission is expected in some regions of the ACT-E region.  

In the rest of this section we describe how each of the components in Eq.~\ref{eqn:spectra_th} are modeled. To allow for comparisons between experiments, we normalize the power spectra at a pivot frequency of $\nu_0=150$~GHz and scale $\ell_0=3000$. This differs slightly from the convention used in previous analyses of the ACT data \citep{fowler/etal:2010,dunkley/etal:2011}. In each case we describe the  parameterization used in the fiducial model; in a later section we consider possible extensions or modifications. 

\subsection{Thermal Sunyaev-Zel'dovich}
\label{subsec:tsz}
\begin{figure}[t]
\includegraphics[scale=0.36,angle=90]{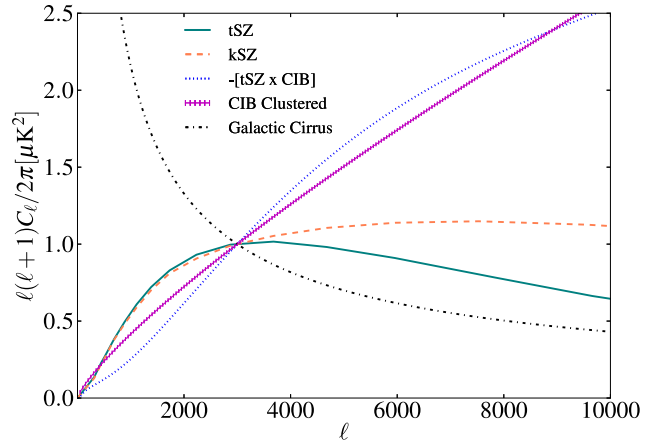}
\caption{Template power spectra for the thermal and kinetic Sunyaev-Zel'dovich effects \citep[tSZ and kSZ,][]{battaglia/etal:2012,battaglia/etal:2010}, clustered CIB sources scaling as $\ell^{0.8}$ \citep[CIB,][]{addison/etal:2012}, the cross-correlation between tSZ and CIB \citep[tSZ-CIB, negative at 150~GHz,][]{addison/dunkley/spergel:2012}, and Galactic cirrus \citep{miville-deschenes/lagache:2005}. They are normalized  at $\ell=3000$ and 150~GHz, and the tSZ-CIB is shown for a perfectly correlated signal. Poisson CIB and radio source power (not shown) scale as $\ell^2$.
\label{fig:templates}
}
\end{figure}

Our model for the power from thermal SZ fluctuations is given by
\be
{\cal B}_\ell^{\rm{tSZ,ij}}  = a_{\rm tSZ}\frac{f(\nu_i)f(\nu_j)}{f^2(\nu_0)}{\cal B}_{0,\ell}^{\rm{tSZ}}, 
\ee
where ${\cal B}_{0,\ell}^{\rm{tSZ}}$ is a template power spectrum corresponding to the predicted tSZ emission at $\nu_0$ for a model with amplitude of matter fluctuations $\sigma_8=0.8$, normalized to $1~\mu$K$^2$ at $\ell_0=3000$, and $a_{\rm tSZ}$ is a free parameter describing its amplitude. An example is shown in Figure \ref{fig:templates}. The factor $f(\nu)=x \coth(x/2)-4$, for  $x=h\nu/k_BT_{\rm CMB}$, scales the expected tSZ emission to thermodynamic units at $\nu$, the effective band-center for the tSZ, given in Table \ref{table:data}. We ignore relativistic corrections \citep[e.g.,][]{itoh/kohyama/nozawa:1998}, since the low-mass clusters that dominate the spectra are well approximated by the non-relativistic formula. This $a_{\rm tSZ}$ normalization differs from that used in \citet{dunkley/etal:2011}. The present choice has the advantage of reducing the dependence on the choice of template, since the main difference between various templates is their amplitude. This means one expects to find the same constraint on $a_{\rm tSZ}$ regardless of template, and a constraint on the SZ power can be converted back into a model-dependent constraint on $\sigma_8$.

The template we adopt is derived from recent hydrodynamic simulations described in \citet{battaglia/etal:2012}. The simulations include the effects of radiative cooling, star formation, and feedback from AGN and supernovae. The predictions are consistent with SZ measurements from both SPT and ACT \citep[e.g.,][]{lueker/etal:2010}, and the shape is shown in Figure \ref{fig:templates}. For the model with $\sigma_8=0.8$, the predicted spectrum reported in \citet{battaglia/etal:2012} has amplitude $a_{\rm tSZ}=5.6\pm0.9$, with standard deviation estimated from ten simulations.

Numerous other authors have also predicted the tSZ spectrum from independent simulations and analytical models \citep[e.g.,][]{komatsu/seljak:2002,shaw/etal:2009,sehgal/etal:2010a,trac/bode/ostriker:2011,shaw/etal:2010,battaglia/etal:2012,efstathiou/migliaccio:2012}, and the expected amplitude for fixed cosmological model varies depending on the astrophysical modeling of the clusters. However, the template shape is broadly consistent among models, and the data are not yet sensitive to shape difference, so we do not include a shape uncertainty. We do not mask clusters, and expect the total SZ power to be the same for ACT and SPT.

\subsection{Kinematic Sunyaev-Zel'dovich}
\label{subsec:ksz}

The kSZ power is expected to have contributions arising from fluctuations in the electron density \citep{ostriker/vishniac:1986}, and in the ionization fraction \citep[e.g.,][]{gruzinov/hu:1998,mcquinn/etal:2005,iliev/etal:2007}, as well as from the motion of galaxy clusters at later times. We model the power as 
\be
{\cal B}_\ell^{\rm{kSZ,ij}}  = a_{\rm kSZ}{\cal B}_{0,\ell}^{\rm{kSZ}},
\ee
where ${\cal B}_{0,\ell}^{\rm kSZ}$ is a template spectrum for the predicted blackbody kSZ emission for a model with $\sigma_8=0.8$, normalized to $1~\mu {\rm K}^2$ at $\ell_0=3000$. The parameter $a_{\rm kSZ}$ describes its normalization. We use a template that assumes a model with instantaneous reionization, described in \citet{battaglia/etal:2010}. This is derived from the same hydrodynamic simulations as the tSZ spectra in Sec \ref{subsec:tsz}, and is shown in Figure \ref{fig:templates}.  
The predicted amplitude from the simulations is $a_{\rm kSZ}=1.5$ for homogeneous reionization at $z=10$ in a $\sigma_8=0.8$ cosmology. This is a quarter of the expected tSZ power. The corresponding kSZ template for the `nonthermal20' model in \citet*{trac/bode/ostriker:2011} has a similar amplitude and shape, as does the \citet*{shaw/rudd/nagai:2012} `CSF' model, and the \citet{bode/etal:prep} model. The power is expected to scale as roughly $\sigma_8^{4.5-5}$ \citep{trac/bode/ostriker:2011,shaw/rudd/nagai:2012}). 

Reionization of the universe is not expected to be instantaneous, as was assumed by \citet{battaglia/etal:2010}. The shape and amplitude of the kSZ power from patchy reionization is far less certain, with simulations predicting a signal at least as large as the homogeneous signal \citep[e.g.,][]{zahn/etal:2012,mesinger/mcquinn/spergel:2012,battaglia/etal:prep}. This gives a total expected signal of $\sim 3$ to $5~\mu {\rm K}^2$ for simple reionization models at $\ell=3000$, comparable to the tSZ at 150~GHz. The dominant effect of patchy reionization on the power spectrum at scales probed by ACT is to alter the amplitude, depending on both the midpoint and duration of reionization \citep[e.g.,][]{battaglia/etal:prep}. We test a modified shape in \S\ref{subsec:extend}, but do not include additional shape uncertainty in the template in the basic model.

\subsection{Cosmic infrared background}
\label{subsec:irpt}

Thermal dust emission from high redshift star-forming galaxies, part of the Cosmic Infrared Background (CIB), is emitted in the rest-frame far infrared and redshifted into the mm-wave range \citep[e.g.,][]{puget/etal:1996,hauser/etal:1998}. Clustering of these galaxies has been detected statistically in mm-wave maps at CMB frequency \citep{hall/etal:2010,dunkley/etal:2011,shirokoff/etal:2011,lagache/etal:2011,hajian/etal:2012,reichardt/etal:2012}, as well as in the sub-mm \citep[e.g.,][]{lagache/etal:2007,viero/etal:2009}. Following the analyses in \citet{dunkley/etal:2011,reichardt/etal:2012,addison/etal:2012}, the power from these galaxies is modeled as the sum of a Poisson and clustered component, given by
\be
{\cal B}_\ell^{\rm{CIB-P,ij}} =a_p \left(\frac{\ell}{\ell_0}\right)^2\left[\frac{\mu(\nu_i,\beta_p)\mu(\nu_j,\beta_p)}{\mu^2(\nu_0,\beta_p)}\right]\mu{\rm K}^2
\ee
for the Poisson part, and 
\be
{\cal B}_\ell^{\rm{CIB-C,ij}}= {a_c}\left(\frac{\ell}{\ell_0}\right)^{2-n}\left[\frac{\mu(\nu_i,\beta_c)\mu(\nu_j,\beta_c)}{\mu^2(\nu_0,\beta_c)}\right] \mu{\rm K}^2
\ee
for the clustered part. Here, $n$ is a power law index, and the frequency scaling of each component is given by a modified blackbody,
\be
\mu(\nu,\beta)=\nu^{\beta}B_{\nu}(T_d)g(\nu),
\ee
with emissivity indices $\beta_p$ and $\beta_c$ for the Poisson and clustered dust terms respectively. The function $B_{\nu}(T_d)$ is the Planck function at frequency $\nu$ for effective dust temperature $T_d$, and the function $g(\nu) = \left(\partial B_\nu(T)/ \partial T \right)^{-1}|_{T_{\rm CMB}}$ converts from flux to thermodynamic units. The parameters $a_p$ and $a_c$ normalize the two components at $\ell_0$ and $\nu_0$, and different frequencies channels are assumed to be perfectly correlated.

The frequency dependence we adopt is an approximation to a sum of modified blackbodies at different redshifts, so this emissivity and temperature are only effective properties of the dust. Following \citet{addison/etal:2012} we fix the dust effective temperature to $T_d=9.7$~K. We also assume $\beta_p=\beta_c$ in the basic model. 

The power-law angular scaling of the clustered term, with increasing $\ell^2 C_\ell$ power at small scales, is shown in Figure \ref{fig:templates}, and approximates the shape of the non-linear power spectrum, which includes contributions from pairs of galaxies in the same dark matter halo, and between galaxies in different halos. \citet{addison/etal:2012} find that a power law in $\ell$ provides a good fit to small-scale power spectra from \planck\ and the Balloon-borne Large-Aperture Submillimeter Telescope (BLAST), and from cross-correlating ACT and BLAST maps. This is consistent with observations of the correlation function from high-redshift Lyman break galaxies, as well as local galaxies \citep[e.g.,][]{giavalisco/etal:1998,connolly/etal:2002}. We fix the power-law index to $n = 1.2$ in the fiducial case ($\ell^{0.8}$), in close agreement with the estimate of $n = 1.25 \pm 0.06$ in \citet{addison/etal:2012}. Both are in agreement with galaxy correlation functions.

We do not expect the CIB power to vary significantly between the ACT and SPT maps, despite the different flux cuts applied to remove sources. Using the model in \citet{addison/dunkley/bond:prep}, the predicted effect of source masking on the CIB power is only at the per cent level.

\subsection{tSZ-CIB cross-correlation}
\label{subsec:cross}

Some spatial correlation is expected between clusters that contribute to the tSZ, and CIB galaxies, since both trace the matter density field. The higher redshift and lower mass groups that make an important contribution to the tSZ signal \citep{komatsu/seljak:2002,trac/bode/ostriker:2011,battaglia/etal:prep} are also likely to host dusty galaxies. \cite{addison/dunkley/spergel:2012} model this correlation, and predict the scale and frequency dependence of its angular power spectrum. For mm-wave spectra at $\ell>2000$, a correlation of $\sim 10$ to $30$\% in power is predicted, with uncertainty dominated by uncertainties in the halo mass and redshift distribution of the CIB. A significant fraction of the CIB power on small scales is due to pairs of galaxies occupying group and cluster-mass halos (the `one-halo' term of the halo model). These same halos are responsible for the tSZ power. A tSZ-CIB correlation in units of power of tens of per cent is therefore possible on small angular scales even if the overall fraction of CIB emission associated with massive halos is small. 

The tSZ-CIB power is negative at 150~GHz, and can partially cancel power from the kSZ effect as it does not vary significantly with frequency over the range probed by ACT and SPT. As a result, neglecting this component can lead to artificially tight constraints on the kSZ power \citep{zahn/etal:2012,mesinger/mcquinn/spergel:2012}. Following \citet{addison/etal:2012} we model the spectrum as
\be
{\cal B}_\ell^{\rm{tSZ-CIB,ij}} =- \xi \sqrt{a_{\rm tSZ}\ a_c}\frac{2f'(\nu_{ij})}{f'(\nu_{0})} {\cal B}_{0,\ell}^{\rm{tSZ-CIB}},
\label{eqn:tsz-cib}
\ee
where ${\cal B}_{0,\ell}^{\rm{tSZ-CIB}}$ is the predicted correlation spectrum shape, normalized to $1~\mu \rm{K}^2$ at $\ell_0$ and shown in Figure \ref{fig:templates}.  The free parameter $\xi$ is the correlation coefficient. The Poisson CIB parameter $a_p$ is not included in Eq.~\ref{eqn:tsz-cib}, unlike in \citet{shirokoff/etal:2011,zahn/etal:2012}, as the sources that dominate the CIB Poisson power in the mm-wave bands are unlikely to have significant redshift-overlap with the tSZ clusters \citep{addison/dunkley/spergel:2012}.
Assuming the same modified blackbody scaling for the CIB as in \S \ref{subsec:irpt}, and tSZ frequency scaling $f(\nu)$, the frequency scaling of the cross-spectra in thermodynamic units is then 
\be
f'(\nu_{ij})=f(\nu_i)\mu(\nu_j,\beta_c)+ f(\nu_j)\mu(\nu_i,\beta_c) ,
\ee
for pivot scale $\nu_0$. 

Since the correlation coefficient is poorly constrained by ACT, we impose a uniform prior of $0<\xi<0.2$ in the basic model; the effect of widening the range to e.g., $\xi<0.5$, corresponding to the maximum allowed correlation in the models explored by \citet{addison/dunkley/spergel:2012}, is discussed in \citet{sievers/etal:prep}. Due to the correlation between $\xi$ and the kSZ power, broadening the limit on $\xi$ increases the upper limit on the kSZ power, but does not affect cosmological results. 

\subsection{Radio point sources}
\label{subsec:radio}

The radio sources at ACT frequencies are not expected to be significantly clustered \citep[see e.g.,][]{sharp/etal:2010,hall/etal:2010}, and to good approximation their power should be perfectly correlated between neighbouring frequencies, consistent with simulations in \citet{sehgal/etal:2010a} and valid for sources with the same spectral indices. As in \citet{dunkley/etal:2011} and \citet{reichardt/etal:2012}, we model the residual power after masking bright sources as Poisson scale-free power, with
\be
 {\cal B}_\ell^{\rm{rad,ij}}  = {a_s} \left(\frac{\ell}{\ell_0}\right)^2\left(\frac{\nu_i\nu_j}{\nu_0^2}\right)^{\alpha_s}\left[\frac{g(\nu_i)g(\nu_j)}{g^2(\nu_0)}\right]\mu{\rm K}^2
\ee
in thermodyamic units, where $g(\nu)$ converts from flux units as for the CIB sources. The amplitude $a_s$ is normalized at $\nu_0$ and $\ell_0$. Measurements of bright sources from ACT and SPT give an estimate for the spectral index in flux units of typically $\alpha_s=-0.5$ \citep{vieira/etal:2010,marriage/etal:2011}. Assuming it holds at fainter fluxes, we fix $\alpha_s=-0.5$ in the fiducial model. 

Bright source counts can also be used to predict $a_s$ by extrapolating to fainter fluxes using a model for the number of sources as a function of flux. This was done in \citet{marriage/etal:2011} for ACT, and for SPT in \citet{keisler/etal:2011,reichardt/etal:2012}. Using point sources measured from the full ACT dataset, \citet{gralla/etal:prep} now predict a residual power $a_s=2.9\pm0.4$ after masking sources brighter than 15 mJy in both ACT regions (the level used to construct the mask for our maps in this analysis), where the catalog is estimated to be complete. We impose this as a Gaussian prior on the power at 150~GHz. For comparison, the estimated power in the SPT power spectra after masking to a flux level of 6~mJy for SPT-high is $a_s = 1.3\pm0.2$ \citep{reichardt/etal:2012}, which we also impose as a Gaussian prior.

\subsection{Residual Galactic cirrus}
\label{subsec:galdust}

The Galactic emission is spatially varying, and \cite{das/etal:prep} show that dust emission can contribute significantly to the power spectra, particularly in our Equatorial region. As reported in \citet{das/etal:prep}, we apply a mask to regions of high dust emission before computing the power spectrum, using measurements at 100~$\mu m$ from IRIS \citep{miville-deschenes/lagache:2005}.

We then marginalize over a residual Galactic cirrus component using a power-law template 
\be
{\cal B}_\ell^{\rm{Gal,ij}}={a_g} \left(\frac{\ell}{\ell_0}\right)^{n_g}\left(\frac{\nu_i\nu_j}{\nu_0^2}\right)^{\beta_g}\left[\frac{g(\nu_i)g(\nu_j)}{g^2(\nu_0)}\right]\mu{\rm K}^2, 
\ee
with amplitude $a_g$, frequency index $\beta_g=3.8$, and angular scaling $n_g=-0.7$. This angular scaling is estimated from the 100~$\mu m$ IRIS dust maps \citep{miville-deschenes/lagache:2005}. The frequency scaling is estimated by correlating the IRIS dust maps with ACT \citep{das/etal:prep}, and is consistent with early results from \planck\  \citep{ade/etal:2011}. Using the correlation coefficients estimated in \citet{das/etal:prep}, we impose priors of $a_{ge}=0.8\pm0.2$, and $a_{gs}=0.4\pm0.2$ in the ACT-E and ACT-S spectra respectively.

For the SPT data, a small Galactic cirrus residual is also expected. In our basic model we follow the treatment in \cite{reichardt/etal:2012}, fixing the Galactic cirrus power to ${\cal B}_{3000}=0.16$, $0.21$, and $2.19$ $\mu {\rm K}^2$ in thermodynamic units at 95, 150, and 220 GHz, with scale dependence $ \propto \ell^{-1.2}$.  However, this model has a steeper angular power law than in our ACT model, and a shallower frequency scaling ($\beta=3.6$ between 150--220~GHz). For consistency we therefore test the effect of adopting the ACT model instead, using $\ell^{-0.7}$ and $\beta_g =3.8$, and a prior of $a_g=0.4\pm0.2$. We find no effect on parameters and no change in the goodness of fit.

\begin{figure*}
\epsscale{0.93} 
\plotone{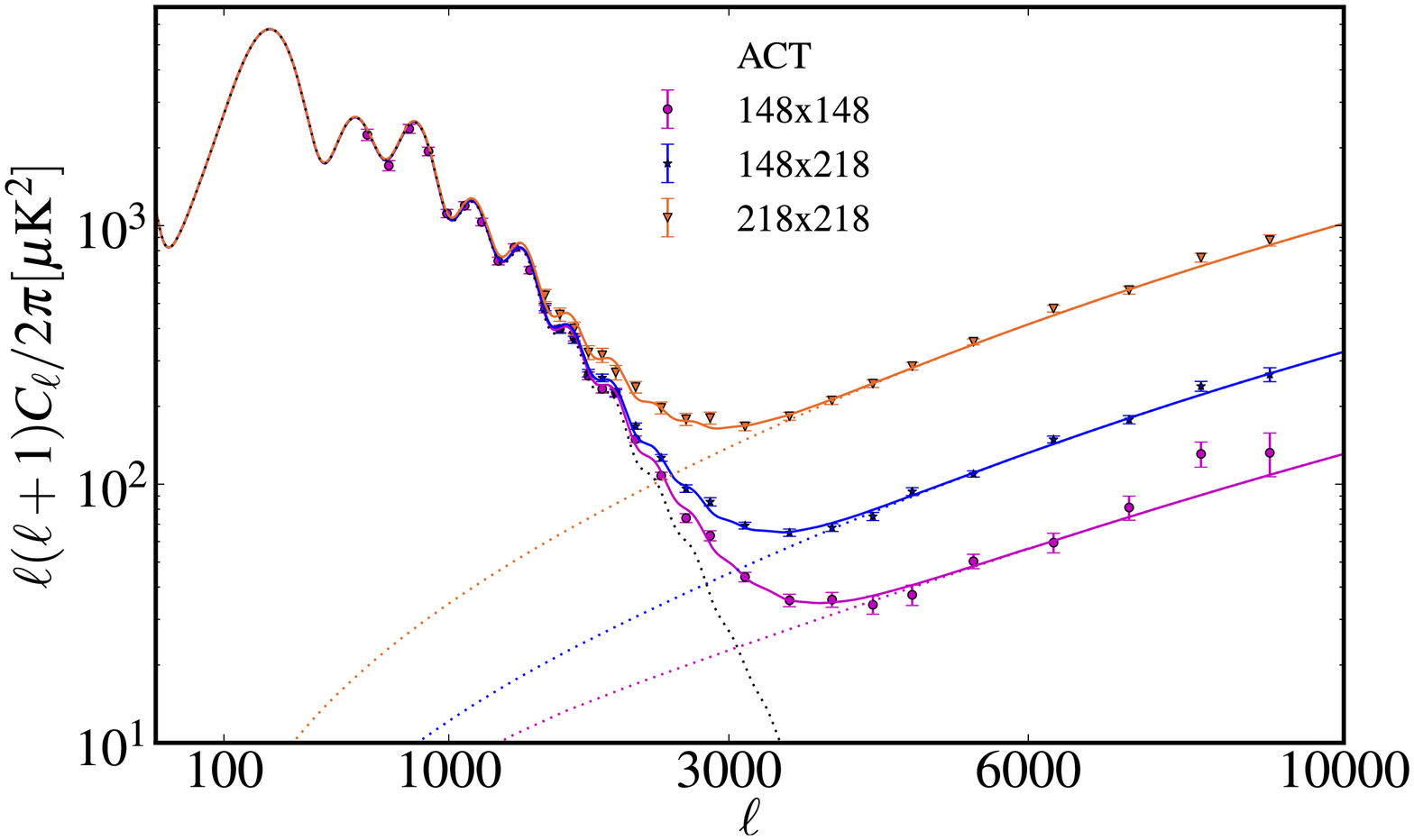}
\epsscale{0.9} 
\plotone{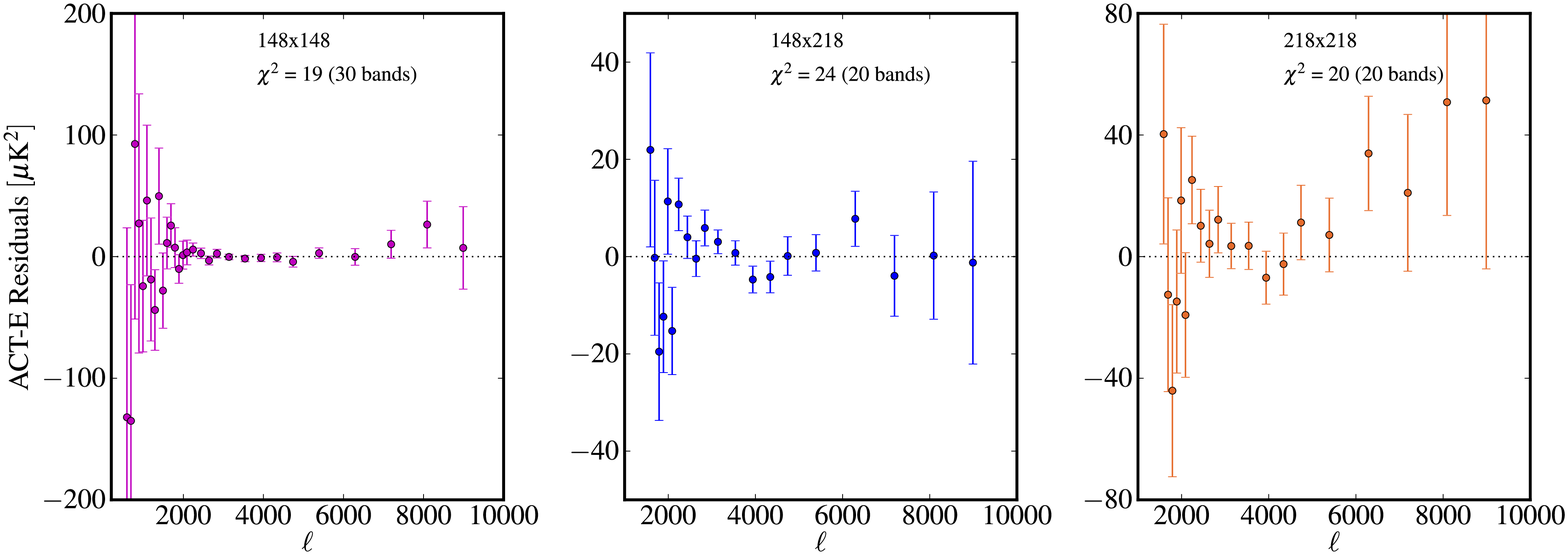}
\vskip-0.5cm
\plotone{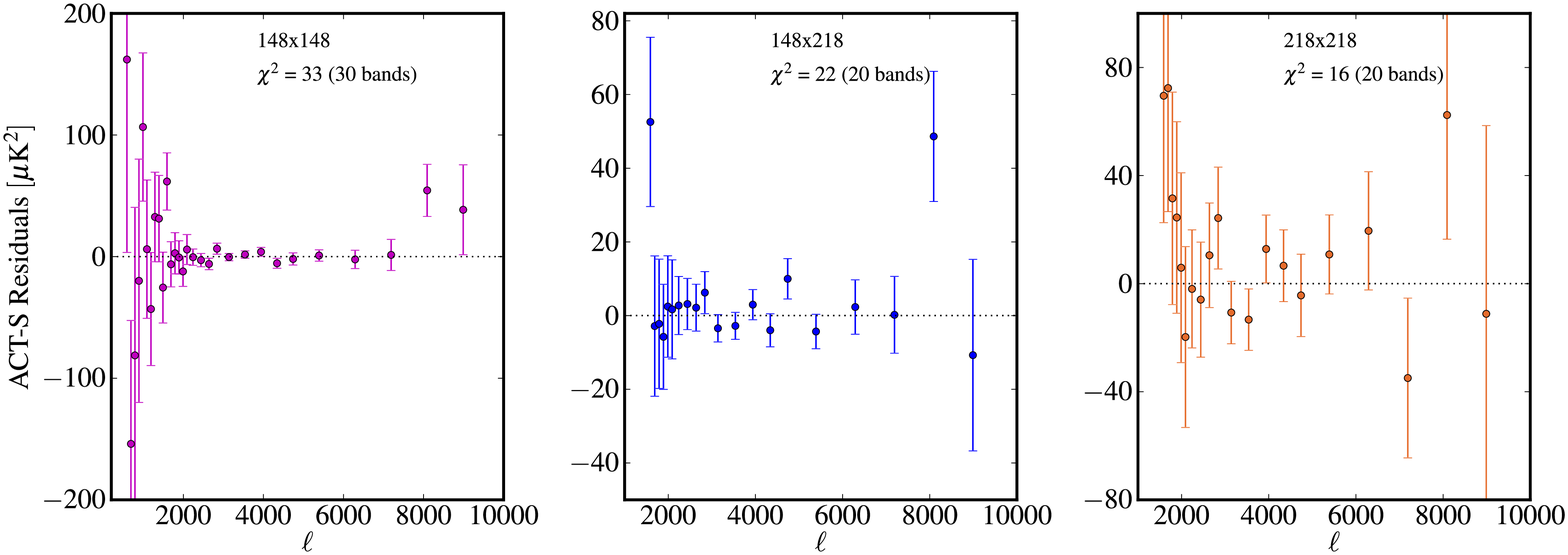}
\caption{(Top) Power spectra measured by ACT \citep{das/etal:prep} at 148 and 218~GHz, and their cross-spectrum, coadded over ACT-E and ACT-S. We show the primary (lensed CMB in dotted black line) and secondary contributions (dotted lines) to the best-fitting model. (Bottom) Residual power in the ACT cross-frequency spectra, after subtracting the best-fitting model, at 148 (left), 148x218 (center), and 218~GHz (right). The errors at small scales are correlated due to beam uncertainty. The model is a good fit simultaneously to ACT-E and ACT-S, with no sigificant residual features.
\label{fig:150_220}
}
\end{figure*}

\section{Full likelihood from small-scale data}
\label{sec:like}

In this section we describe the multi-frequency likelihood used to model the ACT data, and show how we extend it to include other small-scale datasets, in particular data from SPT.

\subsection{Likelihood from the ACT data}
\label{subsec:act_like}

The data from \citet{das/etal:prep} describe the two ACT regions separately (ACT-E and ACT-S); and consist of multi-season and multi-frequency spectra, with an associated covariance matrix\footnote{In this paper we use the original version `v1' of the spectra, unless stated. Since original release, the binning and the beams have been slightly refined, generating `v2' of the spectra. This is described in \citet{das/etal:prep}, and the `v2' spectra are released publicly; cosmological effects are negligible as described in \citet{sievers/etal:prep}.}. They are derived from ACT maps obtained using the method described in \citet{dunner/etal:prep}. The likelihood is Gaussian-distributed to good approximation. To construct the likelihood for each region, given some model spectra $C_\ell^{\rm th,ij}$,  we compute bandpower theoretical spectra using $C_b^{\rm th,ij} = w^{ij}_{b\ell}C_\ell^{\rm th,ij}$, where $w^{ij}_{b\ell}$ is the bandpower window function in band $b$ for cross-spectrum $ij$, described in \citet{das/etal:prep}.

The likelihood, $\mathscr{L}$, of the data for each ACT region separately is given by 
\be
 -2 \ln\mathscr{L} = (C^{\rm th}_b - C_b)^{\rm T} 
 \mat{\Sigma}^{-1}(C^{\rm th}_b -
  C_b) + \ln \det\mat{\Sigma},
\label{eqn:like1}
\ee
where $\Sigma$ is the bandpower covariance matrix. Each of the model and data vectors $C^{\rm th}_b$ and $C_b$ contain three sets of spectra,
\be
C_b = [C_b^{148,148},C_b^{148,218},C_b^{218,218}], \nonumber
\ee
for ACT-E and ACT-S separately, and each spectra set $C_b^{ij}$ itself contains spectra for each cross-season. There are two seasons used for ACT-E (3 cross-season spectra), and three for ACT-S (6 cross-season spectra).  The total likelihood is given by
\be
 -2 \ln\mathscr{L}_{\rm ACT} =  -2 \ln\mathscr{L}_{\rm ACT-E}  -2 \ln\mathscr{L}_{\rm ACT-S}. \nonumber
\label{eqn:like2}
\ee

\subsubsection{Calibration and beam uncertainty}
\label{subsec:calbeamerr}

The data power spectra are calibrated, but have uncertainties. We therefore include a calibration parameter $y_i$, for each map $i$, that scales the estimated data power spectra as
\be
C^{\rm ij}_b \rightarrow y_iy_j {C}^{\rm ij}_b, \nonumber
\label{eqn:cal}
\ee
and the elements of the bandpower covariance matrix as
\be
\mat{\Sigma}^{\rm ij}_{bb'} \rightarrow (y_iy_j)^2\mat{\Sigma}^{\rm ij}_{bb'}. \nonumber
\ee
To account for both ACT regions, we include four calibration parameters: $y_{1e}$, $y_{2e}$ for ACT-E at 148~GHz and 218~GHz, and $y_{1s}$, $y_{2s}$ for ACT-S.

\citet{das/etal:prep} calibrate the \arone\ maps using \wmap, following the method in \citet{hajian/etal:2011}, at an effective $\ell=700$, resulting in a 2\% map calibration error in CMB temperature units. We impose this as a Gaussian prior, with $y_{1e},y_{1s} =1.00\pm0.02$.
The \artwo\ maps are calibrated relative to \arone, at an effective $\ell=1500$. The \artwo\ calibration is constrained by the cross-spectrum, so no prior is imposed on $y_{2e}$ and $y_{2s}$. 
Within each frequency, the individual seasons are calibrated to each other; the  inter-season calibration error is absorbed into the single overall calibration uncertainty. 

 Uncertainties in the measured beam window functions for ACT at 148~GHz are between 0.7 and 0.4\%, and at 218GHz between 1.5 and 0.7\%. We incorporate uncertainties in the measured beams by including them directly in the covariance matrix for the spectra, described in \citet{das/etal:prep}. This technique assumes a fiducial model for the power spectra but is insensitive to its exact form.

\subsubsection{Secondary model parameters}
\label{subsec:sec_param}

Our model described in Section \ref{sec:methods} has nine free secondary parameters for ACT in the basic case: $a_{\rm tSZ}$ and $a_{\rm kSZ}$ describing the SZ emission, $a_p$, $a_c$ and $\beta_c$ describing the CIB power, $a_s$ describing the radio power, $\xi$ describing the tSZ-CIB cross-correlation, and $a_{ge}$ and $a_{gs}$ describing the Galactic cirrus emission. The latter four have strong priors imposed, as described in \S \ref{sec:methods}: $a_s=2.9\pm0.4$, $0<\xi<0.2$, $a_{ge}=0.8\pm0.2$, and $a_{gs}=0.4\pm0.2$. In addition to the nine model parameters, there are four calibration parameters for ACT. 
In \S\ref{subsec:extend} we investigate how additional, or fewer, parameters affect the fit of the model to the data. 
To compute the model requires an effective frequency for each component; we use the band-centers for SZ, radio, and dusty sources given in Table 1 \citep{swetz/etal:2011}. 

\begin{figure*}
\plotone{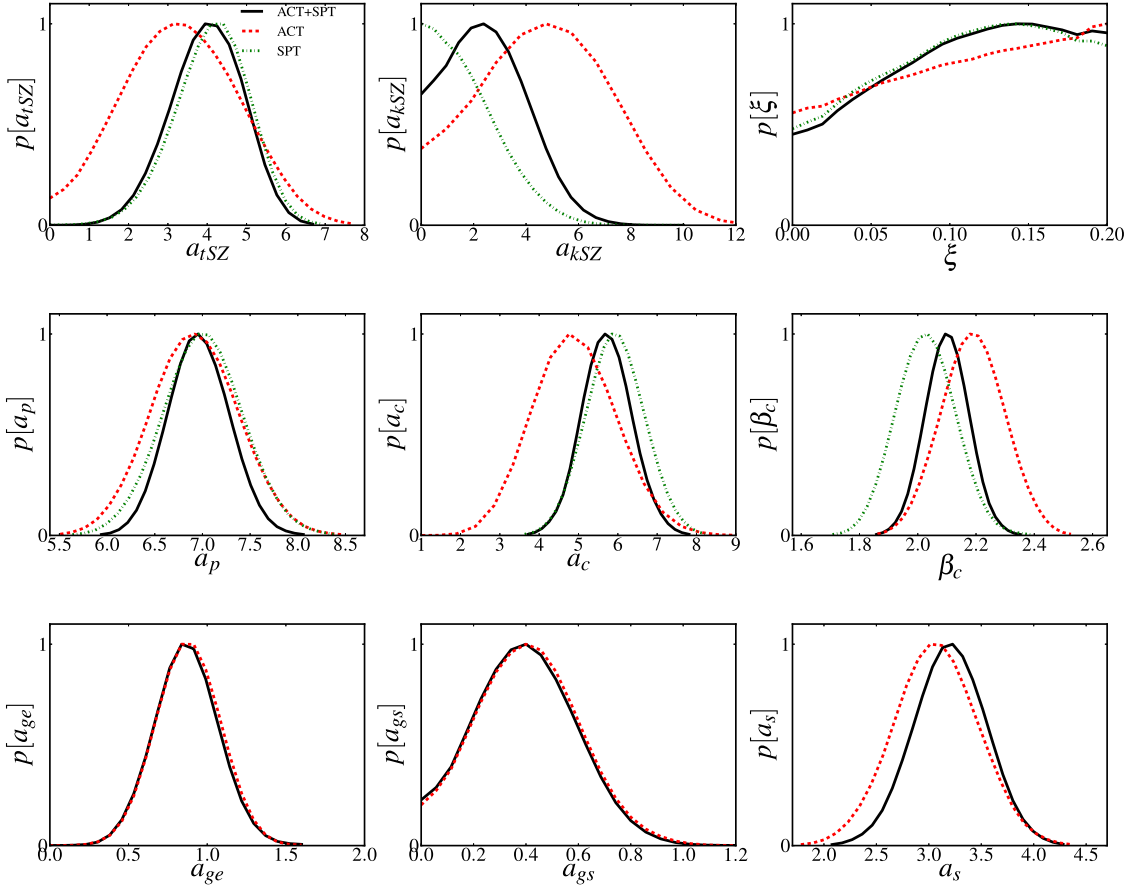}
\caption{Distributions for secondary parameters from ACT and SPT, for best-fitting \LCDM\ model. Parameters \{$a_{\rm tSZ}$, $a_{\rm kSZ}$, $a_p$, $a_c$, $a_{\rm gs}$, $a_{\rm ge}$, $a_s$\} are the $\ell(\ell+1)C_\ell/2\pi$ power in $\mu \rm{K}^2$ at $\ell=3000$ and frequency $150~$GHz. The tSZ-CIB correlation parameter $\xi$ is also defined at $\ell=3000$. The dust emissivity index $\beta_c$ is in flux units, for a modified blackbody with effective temperature $9.7~\rm{K}$. Conversions to power at each frequency are given in Table \ref{table:sec_params}. Strong priors, described in \S\ref{subsec:sec_param}, are imposed on \{$\xi$, $a_{\rm ge}$, $a_{\rm gs}$, $a_s$\}.  \label{fig:sec_params}
}
\end{figure*}

\subsection{Combining with SPT data}
\label{sec:all_like}

The South Pole Telescope observed the sky from 2007--10. Spectra are reported in \citet{keisler/etal:2011} for angular scales $650<\ell<3000$ at 150~GHz, and in \citet{reichardt/etal:2012} for angular scales $2000<\ell<9400$ at 95, 150 and 220~GHz. These observations are summarized in Table \ref{table:data}. One of the goals of our work here is to test for consistency between the two experiments, by using a common framework to describe the SZ and foreground components. As this article was being prepared, refined spectra from SPT at 150 GHz were reported in \citet{story/etal:prep}; we do not include these latest data in our comparison.

Before fitting the SPT data with the ACT secondary model, we confirm that we recover the parameters and $\chi^2$ obtained for the model used in \citet{reichardt/etal:2012} using the SPT data. To combine the data over the full angular range, we follow the method in \citet{reichardt/etal:2012}, using the \citet{keisler/etal:2011} data at $\ell<2000$ (SPT-low) and the \citet{reichardt/etal:2012} data at smaller scales (SPT-high). More radio source power has been removed from the SPT-high spectra due to masking at a deeper flux level, so the expected residual radio power in SPT-low is ${\cal B}_{3000}= 10.5\pm2.4~\mu$K$^2$, compared to $1.3\pm0.2$ for SPT-high. We account for this by first subtracting a radio Poisson power of ${\cal B}_\ell = 9.2~ \mu {\rm K}^2$ from the SPT-low data, following the approach in \citet{reichardt/etal:2012}. A Gaussian prior is then imposed on the overall residual radio level in SPT of $a_{s'} = 1.3\pm0.2$. 

We then extend the ACT secondary model to fit the SPT power spectra. Six of the ACT model parameters are expected to be common for the SPT data (the SZ and CIB parameters: $a_{\rm tSZ}$, $a_{\rm kSZ}$, $\xi$, $a_p$, $a_c$ and $\beta_c$). In addition, to fit the SPT data we require a separate radio source parameter, $a_{s'}$, and three calibration parameters, $y_{1}$, $y_{2}$, $y_{3}$, to calibrate the 95, 150, and 220~GHz maps respectively. We impose a uniform prior on these calibration parameters, as the SPT covariance matrices include the calibration uncertainty.

The likelihood for ACT and SPT together is given by 
\be
-2 \ln\mathscr{L} = -2 \ln\mathscr{L}_{\rm ACT} -2 \ln\mathscr{L}_{\rm SPT} \ .
\ee
The SPT likelihood is constructed as in Eq.~\ref{eqn:like1}, with model and data vectors 
\be
C_b = [C_b^{95,95},C_b^{95,150},C_b^{95,220}, C_b^{150,150},C_b^{150,220},C_b^{220,220}]
\ee
for SPT-high at $\ell>2000$, and $C_b = C_b^{150,150}$ for SPT-low at $\ell<2000$.
To compute the model, we use the band-centers for SZ, radio and dusty sources given in Table 1, from \citet{reichardt/etal:2012}. 

There is some degree of covariance between the ACT and SPT spectra, due to the 54 deg$^2$ overlapping region of sky. Covariance between the two spectra due to cosmic variance, scaling as $1/f_{\rm sky}$, is estimated at a level of 8\% (using $54/\sqrt{590\times800}$); the addition of noise lowers this level, so we neglect the correlation in our combined analysis.

\subsection{Multi-frequency likelihood prescription}
\label{subsec:recipe}

\begin{figure*}
\epsscale{0.7}
\plotone{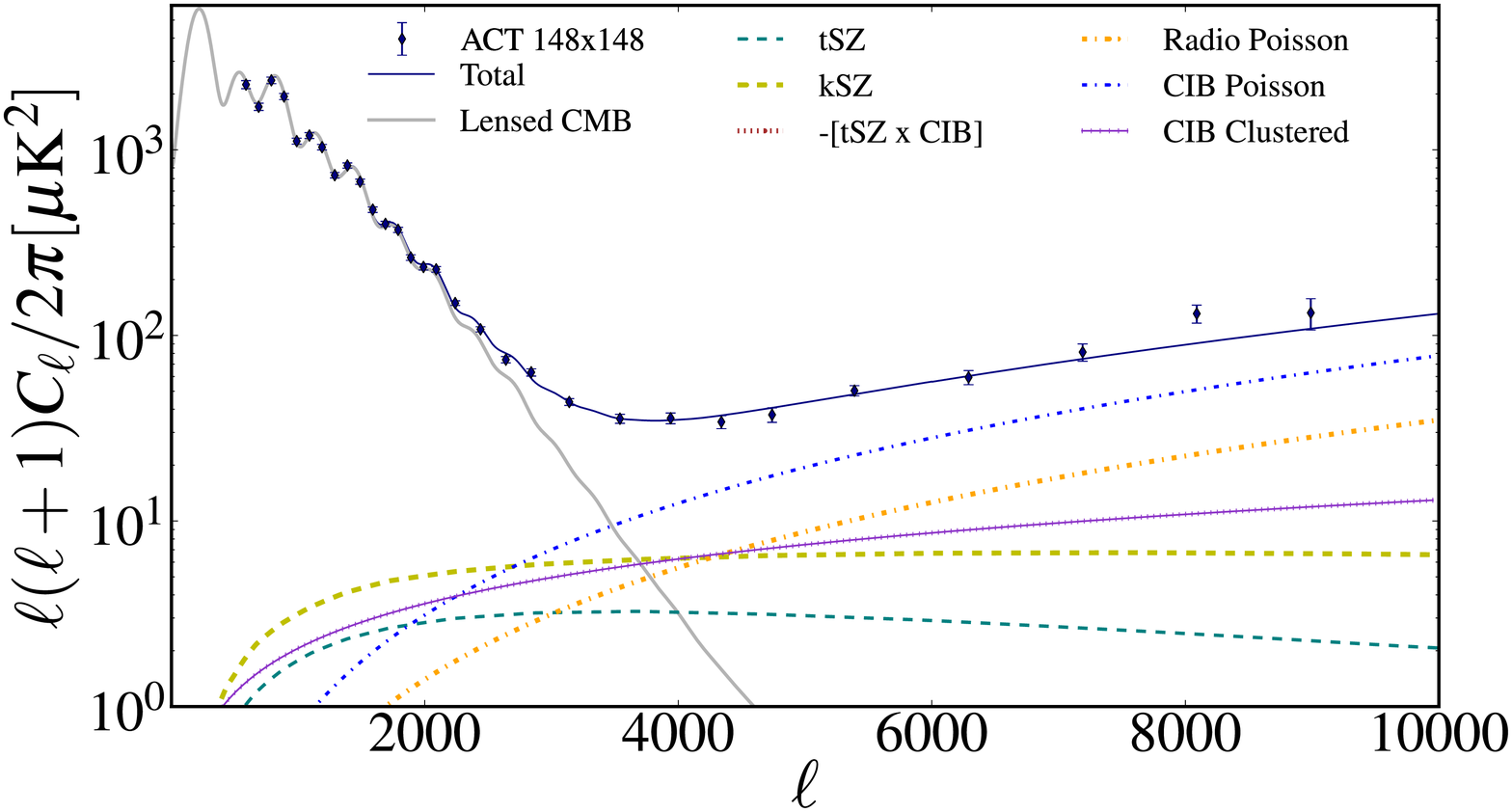}
\plotone{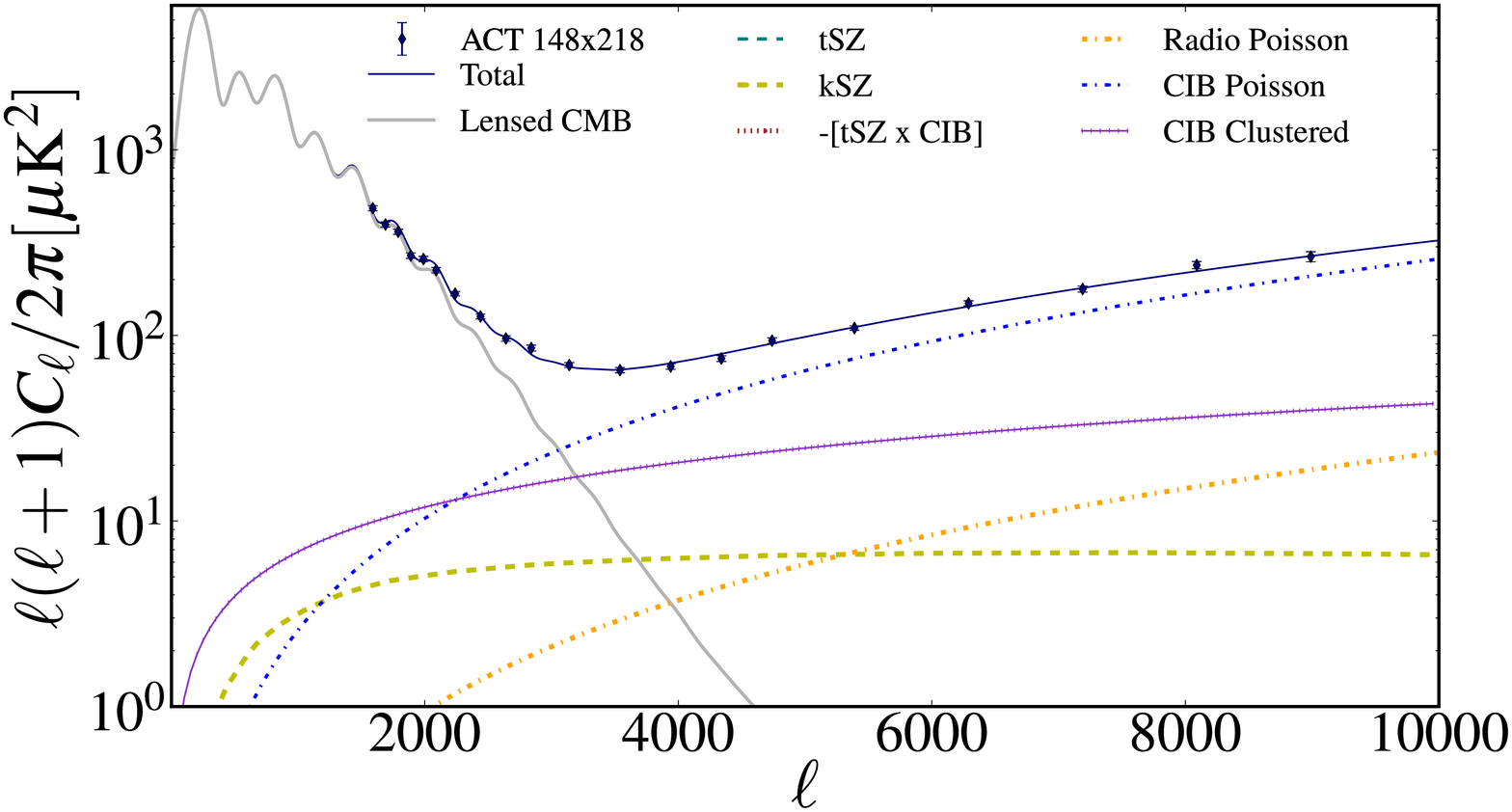}
\plotone{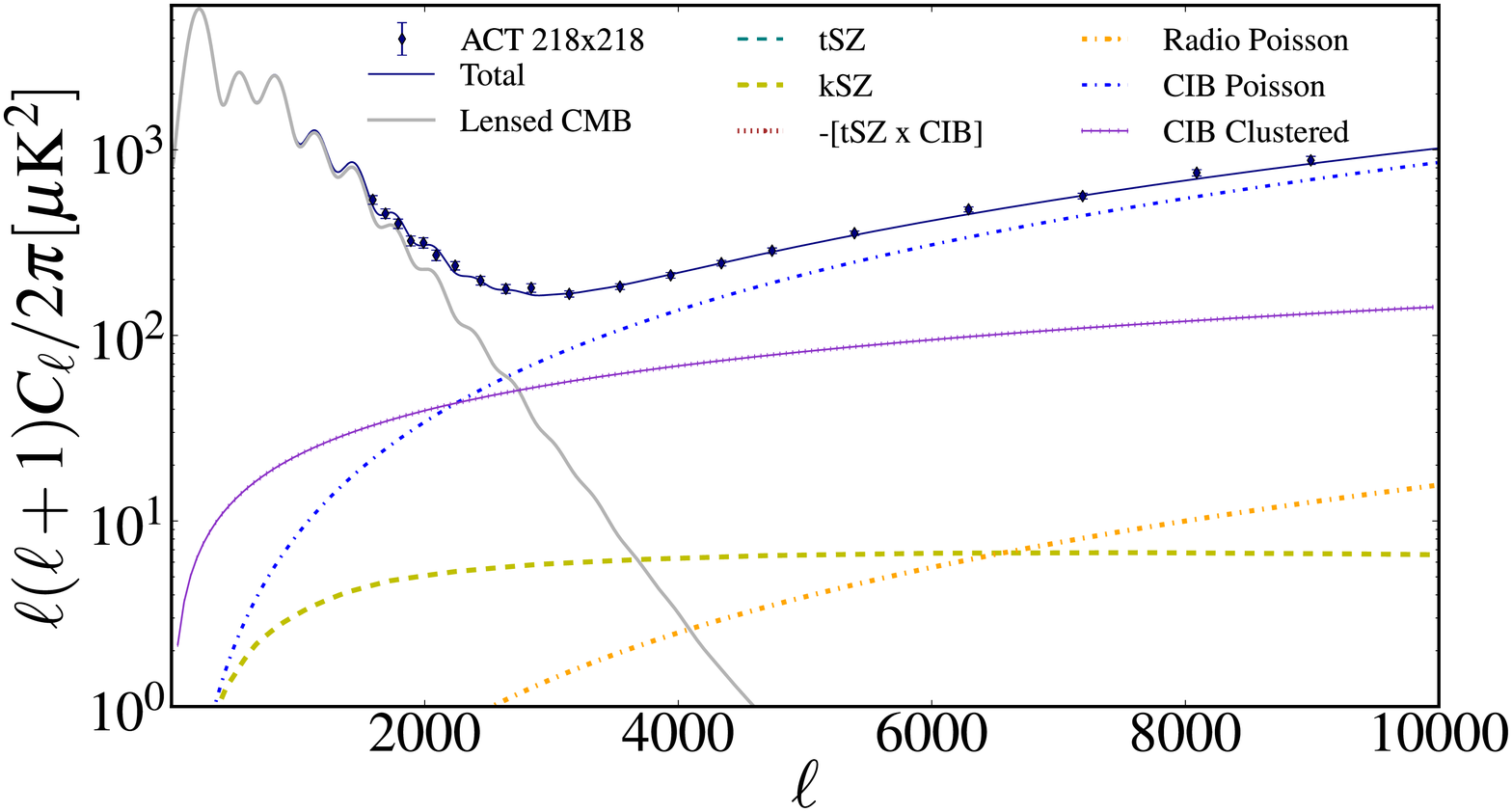}
\caption{Power spectra measured by ACT at 148--218~GHz, with the best-fitting individual SZ and foreground components from Table \ref{table:sec_params}. The Galactic cirrus component has been subtracted. At 148~GHz (top) the secondary components are significant at scales smaller than $\ell\sim2000$, with contributions from tSZ, kSZ, radio galaxies, the CIB, and the tSZ-CIB cross-correlation. The tSZ, kSZ and tSZ-CIB are non-zero in this model, but are not individually significantly detected from the ACT spectra. The radio power is constrained by bright source counts. At 218~GHz (bottom) the secondary signal is significant by $\ell\sim1000$, and is dominated by the Poisson and clustered CIB. 
\label{fig:mult_freq}
}
\end{figure*}

To return the ACT (or ACT+SPT) multi-frequency likelihood for a given model we follow this approach:
\begin{itemize}
\item Select primary cosmological parameters, and compute a theoretical lensed CMB power spectrum ${\cal B}_\ell^{\rm{CMB}}$ using the CAMB numerical Boltzmann code \citep{lewis/challinor/lasenby:2000}. 
\item Select values for common secondary parameters: $\theta=\{a_{\rm tSZ}$, $a_{\rm kSZ}$, $\xi$, $a_p$, $a_c$, $\beta_c$\}.
\item Select values for ACT-specific secondary and calibration parameters: $\theta=\{a_{s}$, $a_{ge}$, $a_{gs}$, $y_{1e}$, $y_{2e}$, $y_{1s}$, $y_{2s}$\}, and/or SPT-specific secondary and calibration parameters: $\theta=\{a_{s'}$, $y_{1}$, $y_{2}$, $y_{3}$\}.
\item Compute the total theoretical secondary power spectra ${\cal B}_\ell^{\rm{sec,ij}}$ for all the required cross-spectra with Eq.~\ref{eqn:spectra_th}, using the effective frequencies for each experiment. 
\item Compute the total model power at each frequency, ${\cal B}_\ell^{\rm{th,ij}}={\cal B}_\ell^{\rm{CMB}}+ {\cal B}_\ell^{\rm{sec,ij}}$.
\item Compute the bandpower theoretical power spectra for each dataset for both South and Equatorial regions for ACT (and for SPT), and compute the likelihood using Eq.~\ref{eqn:like2}.
\end{itemize}

 \subsubsection{Combining with \wmap}
\label{subsec:wmap}

In \citet{sievers/etal:prep} we use the ACT and SPT likelihood in combination with data from \wmap\ seven-year data to estimate cosmological parameters. The \wmap\ data measure $\ell<1000$ angular scales, and so have minimal contamination from SZ and point sources. The public 7-year likelihood estimates the temperature spectrum from V and W bands  \citep[61 and 94~GHz,][]{larson/etal:2011}. At these frequencies and angular scales, the infrared point source contribution is expected to be negligible, consistent with ACT and SPT measurements. The radio point source level is estimated and subtracted internally to the \wmap\ analysis using the multi-frequency data \citep[e.g.,][]{nolta/etal:2009}. Finally, in light of observations by both ACT and SPT, we also neglect the SZ power in the \wmap\ data, as it is expected to be small: $\ell(\ell+1)C_{1000}/2\pi \sim~12~\mu$K$^2$  at 61~GHz from the \citet{battaglia/etal:2012} model, assuming $\sigma_8=0.8$. 

\section{Tests of the multi-frequency likelihood}
\label{sec:results}

In this section we test the goodness of fit of the model to the ACT power spectra, assuming the \LCDM\ cosmological model. We estimate the probability distributions of the secondary parameters using the MCMC method described in \citet{dunkley/etal:2011}, fixing the \LCDM\ parameters at best-fitting values\footnote{Physical baryon density $\Omega_bh^2 =0.02226$, cold dark matter density $\Omega_ch^2=0.1122$, ratio of the acoustic horizon to the angular diameter distance at decoupling $\Theta=1.040$, scalar amplitude $\ln[10^{10} A_s] = 3.186$, spectral index $n_s=0.9707$, and optical depth $\tau=0.898$.}. We then investigate a set of possible extensions or modifications to the secondary model. We include the SPT power spectra and examine the consistency of the foreground model between the two datasets. We use the ACT `v1' spectra for this analysis; as described earlier, a refined estimate of the beams became available after the analysis was complete. We have checked that effects on parameters are negligible ($<0.2\sigma$), so do not update the parameter constraints or plots in this section.

\subsection{ACT data}
\label{subsec:res_act}

We find that the model provides a good fit to the ACT data over the full range of angular scales and frequencies. Figure \ref{fig:150_220} shows the total spectra (coadded over ACT-E and ACT-S, with the best-fitting Galactic cirrus component removed) decomposed into primary and secondary contributions. The SZ and foregrounds dominate at $\ell \gtsim 2400$ at 218~GHz, and at $\ell \gtsim 3200$ at 148~GHz. The goodness of fit is $\chi^2=675$ for $697$ dof (reduced $\chi^2=0.98$, with PTE$=0.72$, for 710 data points and 13 parameters). This indicates a good overall fit, but localised deviations can be hard to identify using the $\chi^2$ over the full angular range. Figure \ref{fig:150_220} therefore shows the residual power after subtracting the best-fitting model; we do not observe any significant features, indicating that the model fits both the angular and frequency dependence of the data in both regions. There is a positive excess in the ACT-E residuals at the smallest scales at 218~GHz, but this is consistent with correlated beam error, accounted for in the covariance matrix. 

The marginalized distributions for the secondary parameters fitting the data are shown in Figure \ref{fig:sec_params} and summarized in Table \ref{table:sec_params}. The Poisson-like and clustered CIB power, $a_p$ and $a_c$, are detected at high significance, with index $\beta=2.2\pm0.1$ consistent with \citet{addison/etal:2012} who find $2.20\pm0.07$. The tSZ and kSZ power are individually seen at low significance, with an anti-correlation between $a_{\rm tSZ}$ and $a_{\rm kSZ}$. The kSZ power peaks at a non-zero value, but the distribution is broad and consistent with zero. The total SZ power is detected at high significance.  The tSZ-CIB correlation coefficient is unconstrained in the prior range $0<\xi<0.2$, and is also unconstrained by ACT if allowed to vary over a broader range (e.g., $\xi<0.5$).  The parameters for the power from radio sources and from Galactic cirrus are driven by their prior distributions. \cite{sievers/etal:prep} present a physical interpretation of these parameters; the constraints are consistent with those found in the 1-year ACT analysis in \citet{dunkley/etal:2011}, with reduced errors.

\begin{figure}
\epsscale{1.3} 
\plotone{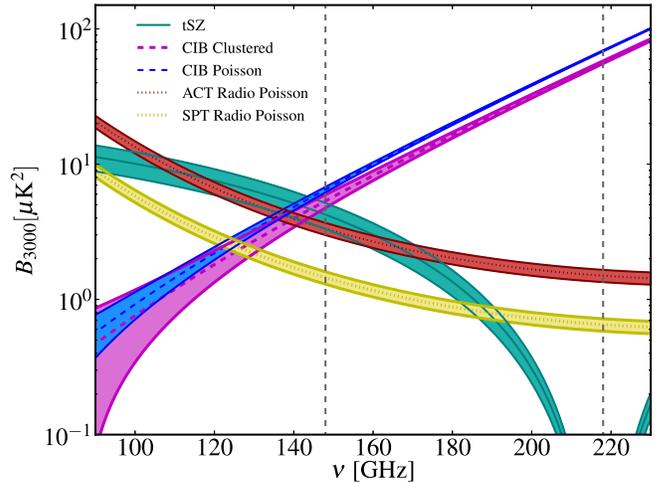}
\caption{Frequency dependence of the dominant components of the foreground power at $\ell = 3000$ measured by the combined ACT and SPT data sets. The bands show the $1\sigma$ uncertainties from Table \ref{table:sec_params}. At $150-220$~GHz the power from fluctuations in the CIB dominates; at lower frequencies the thermal SZ and radio source power is more significant. The SPT radio power is lower due to deeper integration. The kSZ and tSZ-CIB components are not shown. 
\label{fig:for_freq}
}
\end{figure}

In Figure \ref{fig:mult_freq} we show the individual components that contribute to the 148~GHz and 218~GHz power spectra after removal of the best-fitting Galactic cirrus power. At 148~GHz there are contributions from all the components. At 218~GHz the secondary spectrum is dominated by dusty point sources, both clustered and Poisson. This is illustrated further in Figure \ref{fig:for_freq}, which shows the frequency dependence of the dominant components in our model at $\ell = 3000$. The derived constraints on the CIB and radio source components, and the Galactic cirrus emission, at the ACT effective frequencies are also given in Table \ref{table:b3000} to allow comparison with other models. 

\begin{table*}[t]
  \centering
  \caption{ 
    Likelihood parameters, assuming best-fit 6-parameter \LCDM\ for the lensed CMB\tablenotemark{a}.  }
  \begin{tabular}{rccccc}
\hline
&Parameter & Prior\tablenotemark{b} & ACT\tablenotemark{c} & SPT & ACT+SPT\\
\hline 
\hline
SZ &$a_{\rm tSZ}$ & $>0$ & $3.3 \pm 1.4$ & $4.1 \pm 0.9$ & $4.0 \pm 0.9$\\
&$a_{\rm kSZ}$    & $>0$ & $< 8.6$& $< 4.2$ &$ <5.0$\tablenotemark{d} \\ 
&&&&&\\
CIB &$a_{p}$     & $>0$ & $6.9\pm 0.4$& $7.0 \pm 0.4$ &$7.0 \pm 0.3$\\
&$a_{c}$         & $>0$ & $4.9 \pm 0.9$& $6.0 \pm 0.7$ &$5.7 \pm 0.6$\\ 
&$\beta_{c}$     & $>0$ & $2.2\pm 0.1$& $2.0 \pm 0.1$ &$2.10 \pm 0.07$\\ 
&&&&&\\
tSZ-CIB         & $\xi$ &$0<\xi<0.2$& $<0.2$ & $<0.2$ & $<0.2$\\
&&&&&\\
Radio           & $a_{s}$& $2.9\pm0.4$ &$3.1\pm 0.4$ & ---&$3.2 \pm 0.3$\\ 
                & $a_{s'}$ & $1.3 \pm 0.2$ &---& $1.4 \pm 0.1$ &$1.4 \pm 0.1$\\ 
&&&&&\\
Galactic cirrus\tablenotemark{e}          & $a_{ge}$& $0.8 \pm 0.2$& $0.9 \pm 0.2$& ---&$ 0.9 \pm 0.2$\\ 
                & $a_{gs}$& $0.4 \pm 0.2$& $<0.73$&--- &$<0.70$\\ 
&&&&&\\
\hline
Calibration   & $y_{1e}$ & $1.00\pm0.02$ &$1.010\pm 0.007$&--- &$1.006 \pm 0.006$\\
                & $y_{2e}$ & --- &$0.99\pm 0.01$& ---&$0.99 \pm 0.01$\\  
         & $y_{1s}$ & $1.00\pm0.02$ &$1.011\pm 0.007$&--- &$1.010 \pm 0.007$\\
                & $y_{2s}$ & --- &$1.03\pm 0.01$& ---&$1.02 \pm 0.01$\\
            \\
                & $y_{1}$ & --- &---& $1.01 \pm 0.02$ &$1.01 \pm 0.02$\\
                & $y_{2}$ & ---  &---& $1.007 \pm 0.008$ &$1.008 \pm 0.008$\\
                & $y_{3}$ & ---  &---& $1.02 \pm 0.02$ &$1.03 \pm 0.02$\\
\hline
&best fit $\chi^2$/dof  & &  $675/697$& $96/107$ & $773/810$ \\
&PTE  & &  $0.72$& $0.77$ & $0.82$ \\
\hline
     \end{tabular}
\footnotetext[1]{Secondary parameters marginalized over the 6 \LCDM\ model parameters are reported in Table 1 of \citet{sievers/etal:prep}, and are consistent with these results. The marginalization has little effect on these secondary parameters, increasing errors by at most 10\%. }
\footnotetext[2]{A flat prior is imposed, unless indicated as a Gaussian with $x\pm y$ for mean $x$ and standard deviation $y$.} 
\footnotetext[3]{Results are reported as 68\% confidence levels or 95\% upper limits; $\xi$ is unconstrained so the prior upper limit is reported.}
\footnotetext[4]{If the prior on $\xi$  is broadened to $0<\xi<0.5$, the upper limit increases to $a_{\rm kSZ}< 6.9$ \citep{sievers/etal:prep}.}
\footnotetext[5]{The SPT cirrus level we use is ${\cal B}_{3000}=0.16$, $0.21$, and $2.19$ $\mu {\rm K}^2$ at 95, 150, and 220~GHz, as measured in \citet{reichardt/etal:2012}.}
\label{table:sec_params}
\end{table*}

\begin{table}[h]
  \centering
  \caption{Derived constraints on foreground power, ${\cal B}_{3000}~(\mu$K$^2)$ } 
  \begin{tabular}{lcc|ccc}
\hline
& \multicolumn{2}{c|}{ACT} & \multicolumn{3}{c}{SPT} \\
 & 148~GHz  & 218~GHz & 95~GHz & ~150 GHz & ~220GHz \\
\hline
 \hline
CIB-P    & $ 6.8\pm 0.4$  &78 $\pm12$ & $0.90 \pm 0.02$ & $8.0 \pm 0.5$ & $69 \pm 10$\\
CIB-C  & $ 4.8\pm 0.9$ & $54\pm16$ & $0.76 \pm 0.02$ & $6.8 \pm 0.8$ & $59 \pm 12$\\
&& &&& \\
Radio  & $3.2 \pm 0.4$ & $1.4 \pm 0.2$& $7.2 \pm 0.8$ & $1.4 \pm 0.2$ & $0.7 \pm 0.1$\\
&& &&& \\
Gal-E\tablenotemark{a} & $0.9 \pm 0.2$ & $ 11 \pm 2.3$& &  & \\
Gal-S & $0.4 \pm 0.2$ & $ 5.0 \pm 2.3$& &  & \\
\hline
     \end{tabular}
\label{table:b3000}
\footnotetext[1]{Gal-E and Gal-S are the Galactic cirrus powers in the ACT-E and ACT-S spectra. The levels are close to the priors imposed from the measured cross-correlations with IRIS \citep{das/etal:prep}}
\end{table}

\subsection{Combination with SPT}
\label{subsec:res_spt}

\begin{figure}
\epsscale{1.25} 
\plotone{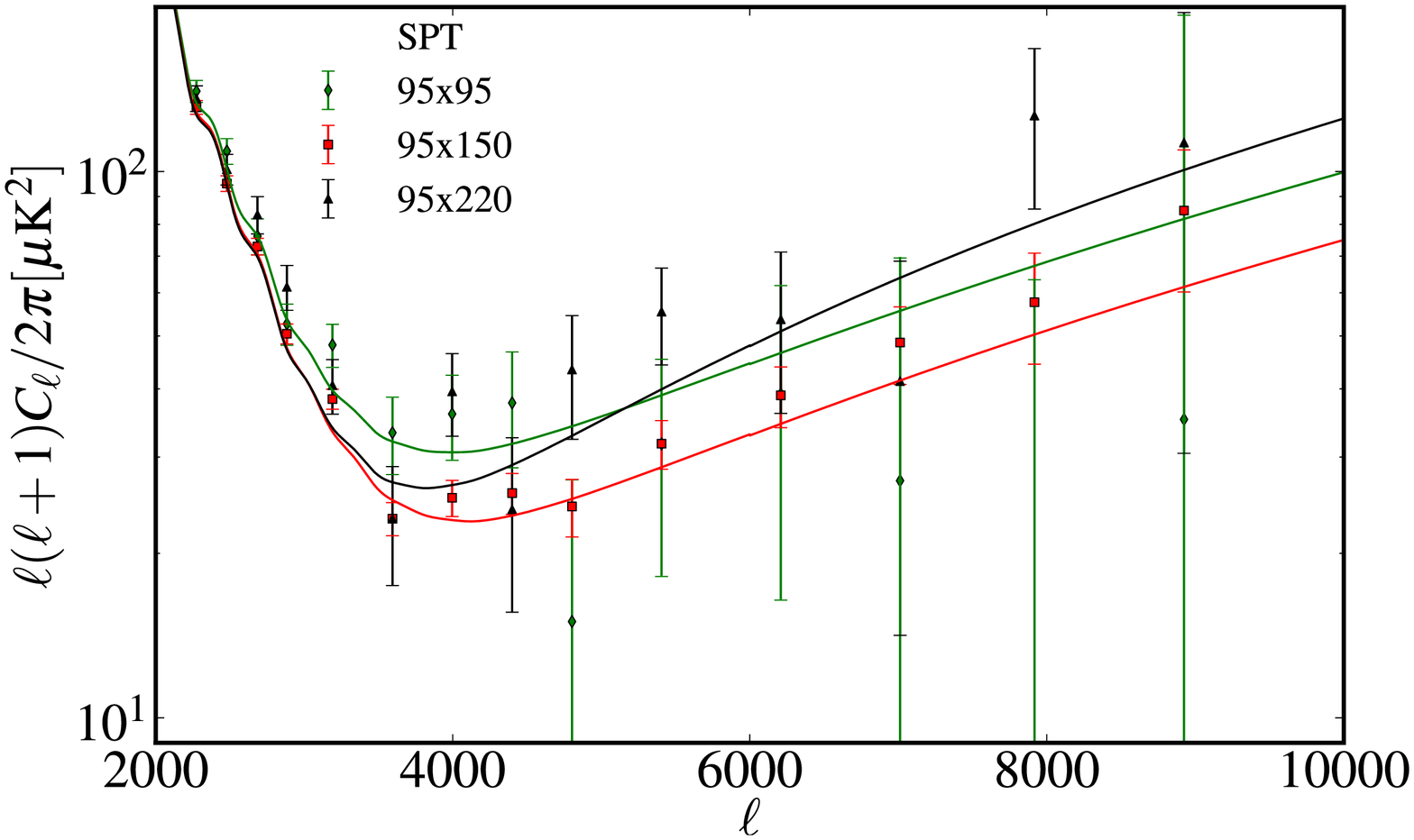}
\plotone{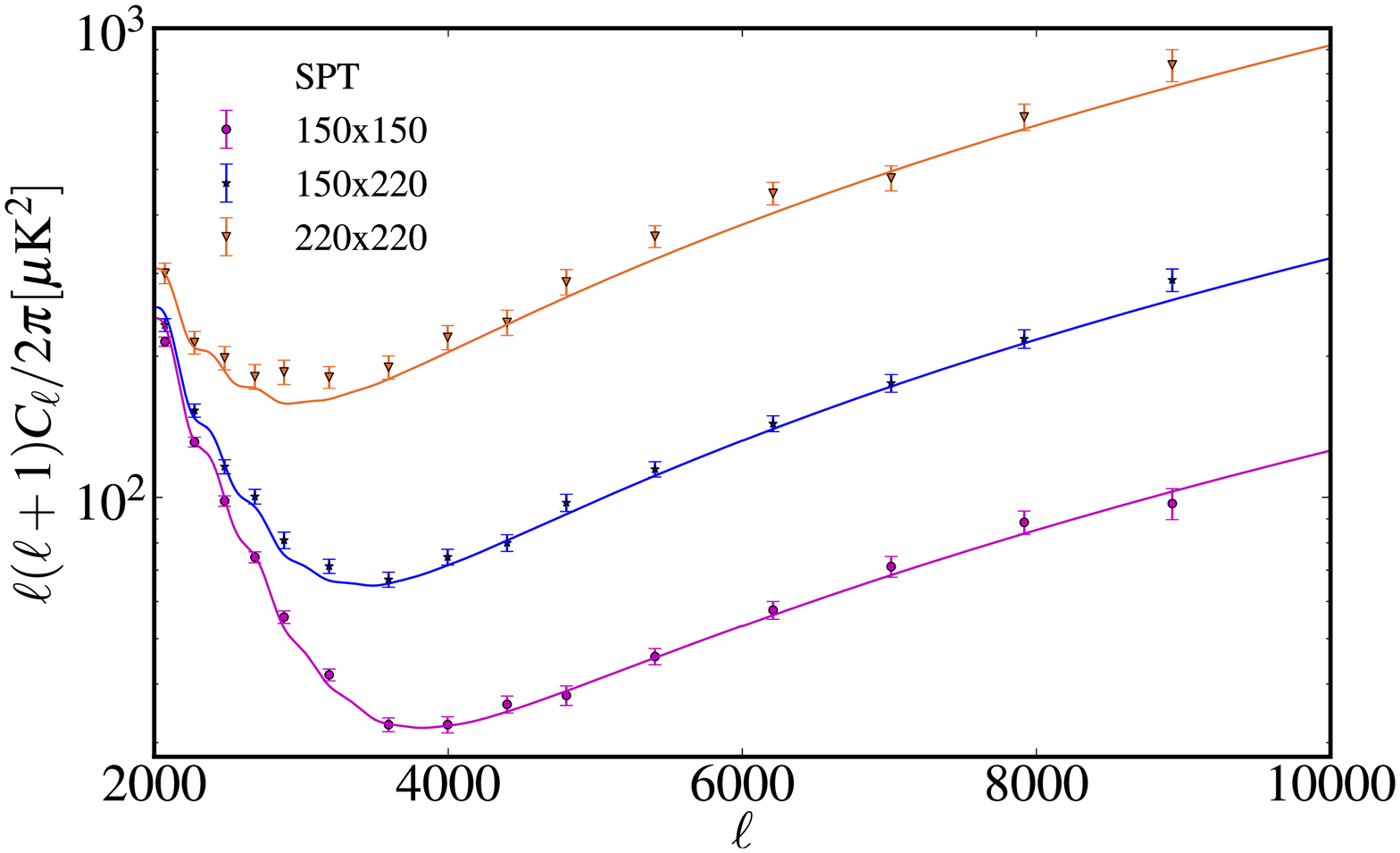}
\caption{Power spectra at  95, 150 and 220~GHz, and their cross-spectra, measured by SPT \citep{reichardt/etal:2012}, fit with the same  model as the ACT data in Figure \ref{fig:150_220}. At 150 GHz the SPT-low spectrum from \citet{keisler/etal:2011} is included, with excess radio power subtracted for comparison. Accounting for the different flux cuts applied to the ACT and SPT maps, and the different bandpass effective frequencies, the spectra are consistent.
\label{fig:spt_spec}
}
\end{figure}

The same model also provides a good fit to the SPT spectra. The SPT data extend the frequency range to 95~GHz, adding three additional cross-spectra to the likelihood. We show the parameters estimated from SPT alone in Figure \ref{fig:sec_params} and in Table \ref{table:sec_params}; they are consistent with those from ACT, with $\sim1\sigma$ shifts in $a_{\rm kSZ}$ and $\beta_c$. The radio Poisson level is lower due to the greater number of radio sources masked in the SPT maps. The model is shown with the SPT spectra in Figure \ref{fig:spt_spec}; the goodness of fit is $\chi^2=96$ for $107$ dof (reduced $\chi^2=0.89$, PTE=0.77). 

Given the consistency of the two datasets, we combine them to generate a joint likelihood; Figure \ref{fig:sec_params} includes the secondary parameters derived from a joint fit. In this case there are ten foreground parameters, and seven calibration parameters. The tenth foreground parameter (not plotted) is $a_{s^\prime}$ for the Poisson radio sources in SPT. The goodness of fit of the joint model is $\chi^2=773$, which can be compared to $\chi^2=675+96 = 771$ for the independent fit to each data set. This supports their consistency.

We report the derived constraints on the CIB and radio source components at the ACT and SPT effective frequencies for each band in Table \ref{table:b3000}. A difference of approximately 15\% is expected between the CIB power at 148~GHz for ACT and 150~GHz for SPT, due to different effective bandpass frequencies and the strong CIB frequency dependence across the mm-wave bands.

\subsection{Tests of the likelihood}
\label{subsec:extend}

\begin{table}
  \centering
  \caption{Modifications to secondary model } 
  \begin{tabular}{lcc}
\hline
Model & Number of  &ACT\tablenotemark{b}  \\
&  parameters\tablenotemark{a} &$\Delta \chi^2$\\
\hline
 \hline
Fiducial       		       & 9  & 0 \\
\\
CIB index $n$ free             & 10  & -4 \\
$\beta_c \ne \beta_p$          & 10  &  0\\

CIB Poisson  corr $= 0.8$& 9 &  5 \\
CIB Clustered  corr $= 0.8$& 9&  1\\
CIB $T_d = 13.6$~K               & 9 & 1\\
\\
Fixed kSZ, $a_{\rm kSZ}=1.5$  & 8 & 3\\
Altered kSZ shape & 9 & 1\\
No tSZ-CIB corr, $\xi=0$                 & 8 & 2 \\
No SZ & 6 & 21 \\
\\
Radio index $\alpha_s = 0$  & 9  & 1\\
\\
No Galactic residual & 7 & 6\\
\hline
     \end{tabular}
\footnotetext[1]{Does not include calibration parameters.} 
\footnotetext[2]{We report the $\Delta \chi^2$ to the nearest integer.} 
\label{table:chisq_extend}
\end{table}

This model fits the ACT and SPT data, and includes our uncertainties about the physical components, with priors describing our knowledge from other observations. However, it is a simplified parameterization of the emission. 
 We therefore consider a set of extensions or modifications to the model, and test how the goodness of fit to the ACT data is affected by an increase or decrease in parameters, or a change in the prior assumptions. In these tests, summarized in Table \ref{table:chisq_extend}, we hold the cosmological model fixed at the best-fitting \LCDM\ parameters. A subset of these extensions are considered further in \citet{sievers/etal:prep}, testing their effect on the primary cosmological parameters.

The CIB appears to be well-fit currently by a power-law in angular scale, with ${\cal B}_\ell \propto \ell^{0.8}$. \citet{addison/etal:2012} find an uncertainty of $0.06$ in this scaling. If we allow the index $n$ to vary, we find no improvement in the fit, but parameter distributions for the CIB parameters $a_p$ and $a_c$ are broadened, as they are correlated with the power-law scaling. We also assume that the CIB emission is perfectly correlated among frequencies, in the range 95--220~GHz. Evidence for imperfect correlation was reported in \citet{lagache/etal:2011}. The effect of this assumption is tested by setting the correlation coefficient to $<1$ in the model, choosing 0.8 for either the Poisson or clustered components, roughly corresponding to the degree of correlation between maps reported in \citet{lagache/etal:2011}. For the Poisson component we find that this degrades the goodness of fit by  $\Delta \chi^2=5$ compared to the perfectly correlated case. We also assume a common frequency scaling of the clustered and Poisson terms. A different scaling in frequency may be expected if, for instance, the redshift dependence of the clustered and Poisson power is different, so a common index is not necessarily expected. Allowing it to vary independently does not significantly improve the goodness of fit, but does lead to a poorly constrained distribution for $\beta_c$, and removes the detection of the CIB at 148~GHz. The index for the Poisson sources in this case, $\beta_p=2.14\pm0.15$, is consistent with the joint index. Changing the effective dust temperature from $9.7$~K to $14$~K, consistent with the value obtained by \citet{gispert/lagache/puget:2000} from a fit to the FIRAS CIB frequency spectrum, has no effect on the model, apart from a corresponding change in $\beta_c$.

Exploring the SZ assumptions, we consider fixing the kSZ contribution to the linear theory estimate for a universe with $\sigma_8=0.8$, assuming homogeneous reionization. This degrades the goodness of fit of the model by only $\Delta \chi^2=3$, indicating that the data cannot yet distinguish between homogenous and patchy reionization. Limiting the kSZ in this way also leads to tighter constraints on the tSZ power. Modifying instead the shape, we find that adding a patchy reionization template from \citet{battaglia/etal:prep} to the kSZ, which changes its shape, does not affect the other secondary parameters. Neglecting the tSZ-CIB cross-correlation also fits the data equally well for one fewer parameter. The dependence of the kSZ constraints on the model for the tSZ-CIB correlation is explored further in \citet{sievers/etal:prep}. If we neglect the SZ components altogether, setting $a_{\rm tSZ}=a_{\rm kSZ}=0$ and keeping only the CIB and Galactic components, the goodness of fit significantly worsens, with an increase of $\Delta \chi^2= 21$.  

Our model imposes an {\it a priori} assumption on the frequency scaling of the extragalactic radio sources. We test the effect of changing the radio spectral index to $\alpha_s=0$, finding negligible effect on parameters and goodness of fit. There is also little effect from changing the prior on the power from \citet{gralla/etal:prep} by 1$\sigma$, corresponding to a different model for the bright radio sources that lie above the detection threshold. Removing the prior altogether opens up degeneracies with other parameters, but does not significantly improve the goodness of fit.

Finally, we test the effect of removing the Galactic cirrus components; the goodness of fit worsens by $\Delta \chi^2=6$ and the clustered CIB level increases. This indicates a preference for Galactic cirrus at the $95\%$ confidence level.

\section{CMB-only likelihood}
\label{sec:cmb}

\begin{figure*}
\epsscale{0.9} 
\plotone{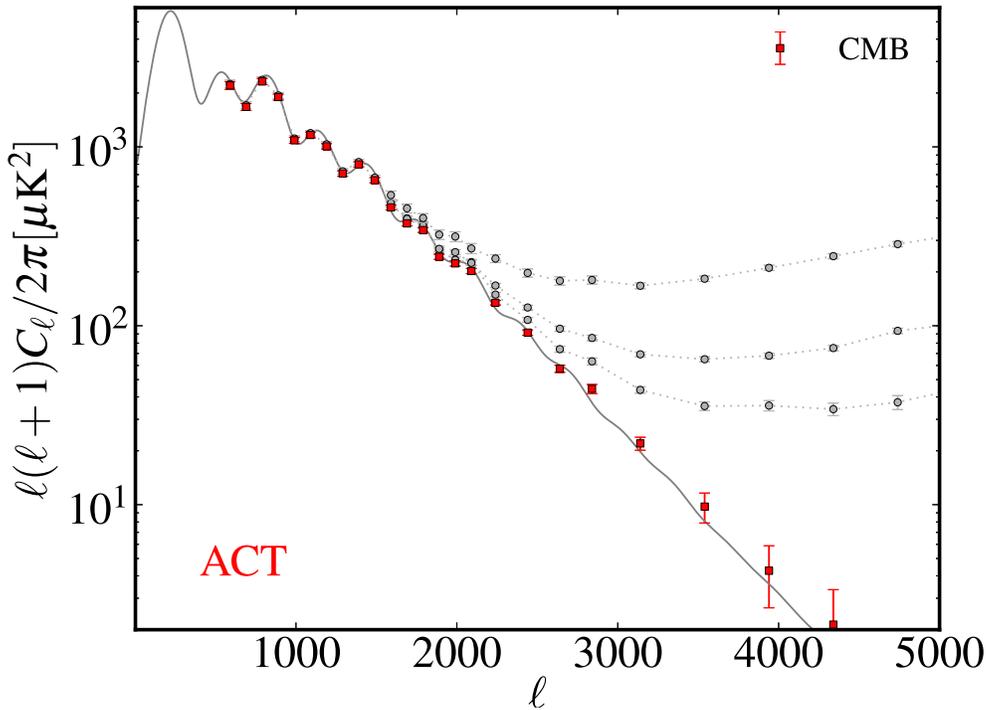}
\caption{Estimated CMB bandpowers from ACT, marginalized over extragalactic source and SZ components. Bandpowers are estimated for ACT-E and ACT-S separately; here we show the inverse-variance weighted combination. The bandpowers are correlated at the $\sim 20\%$ level at scales $\ell\gtsim 2000$ due to covariance with the secondary parameters. The total multi-frequency spectra for ACT-E (dashed, at 148~GHz, 148$\times$218~GHz, and 218~GHz) are also shown to indicate the significant level of SZ and foreground power at small scales.
\label{fig:cmb_only}
}
\end{figure*}

Understanding the contribution of the secondary components to the ACT power spectra is vital for extracting the cosmological information, due to possible degeneracies between primary and secondary parameters. Values of the secondary parameters are also astrophysically interesting. However, if we are only interested in the cosmological parameters, a simplified likelihood is desirable.  

We construct a CMB-only likelihood from ACT data as follows. Instead of using the ACT likelihood to estimate cosmological parameters, we take the intermediate step of estimating the CMB power spectrum in bandpowers, marginalizing over the possible contamination. This is a natural extension to forms of CMB data compression that have been adopted in earlier analyses  \citep*[e.g.,][]{bond/jaffe/knox:2000}. Such `grand unified spectra' were used in a number of subsequent papers to combine the results from various CMB experiments, marginalizing over a variety of nuisance parameters, e.g., in \cite{sievers/etal:2003,bond/contaldi/pogosyan:2003,bond/etal:2004}. At large scales, where contamination from SZ and point sources is negligible, the estimated CMB is simply an optimally combined average of the multi-frequency spectra as in e.g., \citet{hinshaw/etal:2003}. At smaller scales the CMB spectrum has additional uncertainty due to secondary contamination. 

By marginalizing over nuisance parameters in the spectrum-estimation step, we can effectively decouple the primary CMB from non-CMB information. No additional nuisance parameters are then needed when estimating cosmological parameters.

\subsection{Method: bandpowers via Gibbs sampling}

To implement this method in practice, we estimate $n_b$ CMB bandpowers, marginalizing over the secondary parameters. We use the full multi-frequency likelihood from \S \ref{sec:like} to do this, but estimate CMB bandpowers instead of cosmological parameters.

We recall that the model for the theoretical power for a single cross-frequency, cross-season spectrum, $C_\ell^{\rm th,ij}$, is written as
\be
C_\ell^{\rm th,ij}= C_\ell^{\rm CMB} + C_\ell^{\rm sec,ij}(\theta) ,
\ee
where $C_\ell^{\rm sec,ij}(\theta)$ is the secondary signal as in Eq.~\ref{eqn:spectra_th}, and is a function of secondary parameters $\theta$. Writing the spectrum in bandpowers, $C_b^{\rm th,ij} = w_{b\ell}^{ij}C_\ell^{\rm th,ij}$, where $w$ are the bandpower window functions, we write the model for the bandpowers in vector form as 
\be
C_b^{\rm th}= {\bf A} C_b^{\rm CMB} + C_b^{\rm sec}(\theta),
\ee
where $C_b^{\rm th}$ and $C_b^{\rm sec}$ are multi-frequency,  multi-season vectors of length $n_b \times n_{\rm spec}$, where $n_b$ is the number of bandpowers, and $n_{\rm spec}$ is the number of cross-season and cross-frequency spectra ($n_{\rm spec}=9$ for ACT-E, and $18$ for ACT-S). The secondary spectra differ between frequencies but not between seasons. The mapping matrix ${\bf A}$, with elements that are either 1 or 0, maps the CMB bandpower vector (of length $n_b$), which is the same at all frequencies and in all seasons, onto the $(n_b \times n_{\rm spec})$--length data vector. 

We want to estimate $C_b^{\rm CMB}$, marginalized over the secondary parameters, $\theta$. The posterior distribution for $C_b^{\rm CMB}$, given the observed multi-frequency, multi-season spectra $C_b$, can be written as
\be
p(C_b^{\rm CMB}|C_b) = \int p(C_b^{\rm CMB},\theta|C_b) p(\theta) d\theta.
\ee
We find that Gibbs sampling provides an efficient way to map out the joint distribution $p(C_b^{\rm CMB},\theta|C_b)$, and to extract the desired marginalized distribution $p(C_b^{\rm CMB}|C_b)$.

 Gibbs sampling can be used in the special case that at least one conditional slice through a multi-dimensional distribution has a known form, and has been used, for example, to estimate the large-scale CMB power spectrum, and to marginalize over Galactic foregrounds \citep[e.g.,][]{wandelt/larson/lakshminarayanan:2004,jewell/levin/anderson:2004,eriksen/etal:2004,dunkley/etal:2009,larson/etal:2011}. Here, we split the joint distribution into two conditional distributions: $p(C^{\rm CMB}_b|\theta,C_b)$, and  $p(\theta|C^{\rm CMB}_b,C_b)$.
We write the multi-frequency likelihood for a single ACT region, from Eq.~\ref{eqn:like1}, as 
\ba
-2 \ln\mathscr{L} &=& ({\bf A}C_b^{\rm CMB}+C_b^{\rm sec} - C_b)^{\rm T} 
 \mat{\Sigma}^{-1}({\bf A}C_b^{\rm CMB}+C_b^{\rm sec} -  C_b) \nonumber \\
&&+ \ln \det\mat{\Sigma},
\label{eqn:like_gibbs}
\ea
which is a multivariate Gaussian. 

If $C_b^{\rm sec}$ is held fixed, the conditional distribution for the CMB bandpowers, $p(C_b^{\rm CMB}|\theta,C_b)$,  assuming a uniform prior for $p(C^{\rm CMB}_b)$, is then also a Gaussian. It has a distribution given by
\ba
 -2 \ln p(C^{\rm CMB}_b|\theta,C_b) &=& (C_b^{\rm CMB}-{\hat C}_b)^{\rm T}\mat{Q}^{-1} (C_b^{\rm CMB}-{\hat C}_b)\nonumber \\
&&+ \ln \det\mat{Q},
\ea
The mean, $\hat C_b$, and covariance, ${\mat Q}$, of this conditional distribution are obtained by taking the derivatives of the likelihood in Eq. \ref{eqn:like_gibbs} with respect to $C_b^{\rm CMB}$. This gives mean
\be
{\hat  C}_b= [{\mat A}^T{\mat \Sigma}^{-1}{\mat A}]^{-1}[{\mat A}^T{\mat \Sigma^{-1}}(C_b - C_b^{\rm sec})], 
\ee
and covariance
\be
{\mat Q}={\mat A}^T{\mat \Sigma}^{-1}{\mat A}.
\ee
We can draw a random sample from this Gaussian distribution by taking the Cholesky decomposition of the covariance matrix, ${\bf Q}= {\bf L} {\bf L}^T$, and drawing a vector of Gaussian random variates $G$. The sample is then given by 
$C^{\rm CMB}_{b}= {\hat C}_b + {\bf L} G$. 

If instead $C^{\rm CMB}_b$ is held fixed, the conditional distribution for the secondary parameters, $p(\theta|C^{\rm CMB}_b,C_b)$, is not a Gaussian, but can be sampled with the Metropolis algorithm that is used in the MCMC sampling in \S \ref{sec:results}. To map out the full joint distribution for $\theta$ and $C^{\rm CMB}_b$ we alternate a Gibbs sampling step, drawing a new vector of CMB bandpowers, $C^{\rm CMB}_b$, with a Metropolis step, drawing a trial vector of the secondary parameters $\theta$. 

We choose a uniform positive prior distribution for $p(C^{\rm CMB}_b)$, and restrict the CMB bandpowers to be zero at $\ell>4500$, where the CMB power is expected to be less than 1 $\mu$K$^2$. About 100,000 steps are required for convergence of the joint distribution, assessed with the \citet{dunkley/etal:2005} spectral test. The mean and covariance of the resulting marginalized bandpowers, $C^{\rm CMB}_b$, are then estimated following the standard MCMC prescription \citep[e.g.,][]{lewis/bridle:2002,spergel/etal:2003}.

\subsubsection{Combining spectra from different regions}

There is only one underlying CMB power spectrum, so this method could be used to estimate a single spectrum, or set of bandpowers, from the two ACT regions. However, the bandpower window functions are different for each region due to their distinct geometries. To easily conserve this information, we estimate the CMB bandpowers for ACT-E and ACT-S separately. Since the secondary parameters are common to both, the estimated CMB bandpowers will be correlated between the regions at small scales. 

To estimate the joint distribution for the ACT-E and ACT-S bandpowers, we map out $p(C_b^{\rm CMB-E},C_b^{\rm CMB-S}, \theta|C_b)$ by taking sequential sampling steps from the conditional distributions:
\ba
p(C_b^{\rm CMB-E}|C_b^{\rm CMB-S},  \theta, C_b), \nonumber \\
p(C_b^{\rm CMB-S}| C_b^{\rm CMB-E},  \theta, C_b), \nonumber \\
p( \theta| C_b^{\rm CMB-E},C_b^{\rm CMB-S}, C_b). 
\ea 
The marginalized distribution for the CMB bandpowers, $p(C_b^{\rm CMB-E},C_b^{\rm CMB-S}|C_b)$, with its associated covariance matrix, is then computed from the samples. This could be extended to include the SPT data, or data from \planck, for example.

\subsubsection{Calibration factors}

There are four ACT calibration factors. To minimize bin-to-bin correlations in the estimated CMB bandpowers due to calibration uncertainty, we divide out the 148~GHz calibrations for the two ACT spectra, estimating $C_b'^{\rm CMB-E} = C_b^{\rm CMB-E}/y^2_{1e}$ for the ACT-E bandpowers, and $C_b'^{\rm CMB-S} = C_b^{\rm CMB-S}/y^2_{1s}$ for ACT-S. 

We then estimate the 148~GHz calibration factors, $y_{1e}$, $y_{1s}$, and relative 218/148 GHz calibration factors, $y_{2e}/y_{1e}$, $y_{2s}/y_{1s}$, as part of the secondary parameter set.

\subsection{Marginalized CMB bandpowers}
\label{subsec:bands}

Figure \ref{fig:cmb_only} shows the estimated CMB bandpowers from the ACT-E and ACT-S spectra, co-added together and compared to the multi-frequency spectra. The bandpowers for each region are reported in Table \ref{table:cls}. In this table we report the CMB spectra derived using the updated `v2' multi-frequency spectra. Without assuming any cosmological model, the CMB bandpowers over the full angular range are remarkably consistent with the theoretical \LCDM\ model predicted by \wmap. The uncertainty on the bandpowers rises at scales smaller than $\ell\sim3000$, and the correlations between bandpowers increases. 

\begin{figure}
\epsscale{1.2} 
\plotone{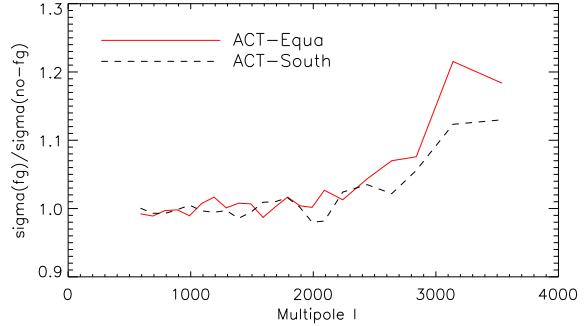}
\caption{Inflation of errors due to foreground marginalization, relative to the errors for a best-fitting foreground model: at scales smaller than $\ell\sim 2000$, the errors are increased due to foreground uncertainty.
\label{fig:err_ratio}
}
\end{figure}

Figure \ref{fig:err_ratio} shows the effect of marginalization on the bandpower errors, using the ratio between the marginalized errors and the unmarginalized errors for a fixed secondary model. It is clear that by measuring the spectrum at multiple frequencies, the CMB can be successfully separated from secondary contamination out to scales $\ell \ltsim 3500$. At scales $\ell<2000$ there is little error inflation due to foreground uncertainty, and the errors are inflated by $\sim$20\% (15\%) by $\ell=3200$ for the ACT-E (ACT-S) spectra.  The marginalized distributions for the CMB bandpowers are well approximated by Gaussians for multipoles to band-center $\ell=3540$ for ACT, as shown in the Appendix. 

We compare the secondary parameters recovered in this model-independent sampling to the case where \LCDM\ is assumed. This comparison is shown in the Appendix; the parameters are consistent, with about a $1\sigma$ shift in the estimated kSZ power. We find that the CMB bandpowers are not strongly correlated with the secondary parameters until scales well into the Silk damping tail at $\ell\gtsim2500$; a dominant correlation is then with the kSZ power due to its blackbody frequency dependence, and a smaller kSZ power -- compensated by larger primary CMB power -- is allowed when the \LCDM\ assumption is relaxed. The CMB bandpower covariance matrix conserves this correlation information.

\begin{table}[h]
  \centering
  \caption{ \label{table:cls}
Lensed CMB anisotropy power\tablenotemark{a}
  }
  \begin{tabular}{rrrr}
 $\ell_b$  & \multicolumn{3}{c}{$\ell(\ell+1)C_b/2\pi$ $(\mu$K$^2$)} \\
 & ACT-E & ACT-S & Coadd\tablenotemark{b}\\
\hline
590 & $2157\pm 159$ & $2343\pm160$ & $2250\pm113$  \\
690 & $1729\pm 115$ & $1744\pm107$ & $1737\pm78$   \\
790 & $ 2499\pm 146$ & $2274\pm126$ & $2370\pm96$  \\
890 & $ 1945\pm 109$ & $1903\pm102$ & $1923\pm74$  \\
990 & $ 1068\pm 58$ & $1187\pm61$ & $1124\pm42$  \\
1090 & $ 1206\pm 61$ & $1149\pm58$ & $1176\pm42$  \\
1190 & $ 1036\pm  51$ & $1016\pm48$  & $1026\pm35$  \\
1290 & $ 679\pm  33$ & $766\pm38$ & $717\pm25$  \\
1390 & $ 819\pm  39$ & $787\pm36$ & $802\pm26$  \\
1490 & $ 661\pm 31$ & $650\pm29$ & $655\pm21$  \\
1590 & $ 452\pm 19$ & $474\pm21$ & $462\pm14$ \\
1690 & $ 387\pm 16$ & $355\pm17$ & $372\pm11$  \\
1790 & $ 344\pm 14$ & $347\pm16$ & $345\pm10$  \\
1890 & $ 242\pm 10$ & $246\pm12$ & $244\pm8$  \\
1990 & $ 230\pm 10$ & $214\pm11$ & $223\pm8$  \\
2090 & $ 199\pm 9$ & $204\pm11$  & $201\pm7$  \\
2240 & $ 137\pm 5$ & $127\pm6$   & $133\pm4$  \\
2440 & $ 92.9\pm 3.9$ & $87.8\pm5.1$ & $91.0\pm3.1$  \\
2640 & $ 57.6\pm 3.3$ & $56.4\pm4.5$ & $57.2\pm2.6$  \\
2840 & $ 43.0\pm 3.2$ & $44.7\pm4.2$ & $43.6\pm2.5$  \\
3140 & $ 22.5\pm 2.2$ & $19.3\pm2.8$ & $21.3\pm1.7$  \\
3540 & $ 9.2\pm2.3$ & $9.0\pm2.8$ & $9.1\pm1.8$  \\
\hline
\footnotetext[1]{To compute a likelihood using these data, ACT-E and ACT-S should be used with the covariance matrix and bandpower window functions provided on LAMBDA.}
\footnotetext[2]{This coadds the ACT-E and ACT-S CMB bandpowers for plotting purposes.}
\end{tabular}
\end{table}

\subsection{The CMB-only likelihood}

We construct the CMB-only likelihood from the angular range where the CMB bandpowers are Gaussian, conservatively choosing $\ell<3500$. We do not use the $3500<\ell<4500$ bandpowers as they are increasingly non-Gaussian, due to the foreground marginalization, and are more strongly correlated with foreground parameters. The likelihood is given by
\ba
 -2 \ln\mathscr{L}({\tilde C}_b^{\rm CMB}|C_\ell^{\rm th}) =  x^{\rm T}  \mat{{\tilde \Sigma}}^{-1}x  + \ln \det\mat{ {\tilde \Sigma}}.
\label{eqn:cmblike}
\ea
Here
\be
x = \left(\begin{array}{c} {\tilde C}^{\rm CMB-E}_b-w_{b \ell, \rm ACT-E} C_\ell^{\rm th}  \\ {\tilde C}^{\rm CMB-S}_b-w_{b \ell, \rm ACT-s} C_\ell^{\rm th}  \end{array}\right),
\ee
where ${\tilde C}^{\rm CMB}_b$ and $\mat{{\tilde \Sigma}}$ are the marginalized mean and covariance matrix for the bandpowers, and $C_\ell^{\rm th}$ is the lensed CMB spectrum generated from e.g., CAMB.  We use 21 bandpowers for ACT-E and ACT-S in the range $500<\ell<3500$. A single calibration parameter for each region is marginalized over analytically, following \citet{bridle/etal:2002}. The prescription for using this likelihood is simple, as no extra nuisance parameters are needed.

To test the performance of this compressed likelihood, results are compared using the full multi-frequency likelihood, and the CMB-only likelihood. Cosmological parameters are estimated for the restricted \LCDM\ 6-parameter model, and a set of more extended models that probe the damping tail and peak shapes, including the running of the spectral index, the number of relativistic degrees of freedom $N_{\rm eff}$, the lensing amplitude $A_L$, and the variation in fine structure constant, $\alpha$. Parameter constraints using both likelihoods agree to $0.1\sigma$, and are reported in  \citet{sievers/etal:prep}.
We conclude that this is an efficient alternative to the full likelihood for the typical extensions considered in cosmological analyses, although the full likelihood may give more optimal results for unusual models with features far into the damping tail.

\section{Summary}
\label{sec:conclude}

In this paper we have presented a likelihood formalism to describe the ACT multi-frequency power spectra that includes contributions from SZ and foreground components in addition to the lensed CMB. We model the data including four late-time astrophysical components: thermal and kinetic SZ, emission from CIB galaxies, and emission from radio galaxies. 

We have quantified these components using seven power spectra, splitting the CIB into a Poisson and clustered part, and including power from the cross-correlation between tSZ emission from clusters, and emission from CIB galaxies that also trace the large scale structure. Rather than a minimal model with the fewest parameters, we have sought a model that includes our uncertainties with priors describing our knowledge from additional observations. For example, while the data do not demand that we include the tSZ-CIB correlation, we are motivated to include it to avoid placing unphysically strong limits on the kSZ power.

Modeling these astrophysical components allows us to probe the primordial CMB fluctuations down to an angular resolution of 4' using ACT. We have used the model to extract an estimate of the primordial CMB spectrum well into the Silk damping tail, marginalizing over the foreground uncertainty. This produces a simplified compressed likelihood for use in cosmological parameter estimation.

We find that data observed by the South Pole Telescope give results consistent with ACT, accounting for the different removal of radio point sources, and different degree of contamination by Galactic cirrus.  SPT and ACT have very different instrument design and scan strategies, and their observations on the sky have limited overlap. The excellent agreement between the datasets is not only an important cross-check but is another demonstration of cosmic homogeneity.

\acknowledgements

This work was supported by the U.S. National Science Foundation through awards AST-0408698 and AST-0965625 for the ACT project, as well as awards PHY-0855887 and PHY-1214379. Funding was also provided by Princeton University, the University of Pennsylvania, and a Canada Foundation for Innovation (CFI) award to UBC. ACT operates in the Parque Astron\'omico Atacama in northern Chile under the auspices of the Comisi\'on Nacional de Investigaci\'on Cient\'ifica y Tecnol\'ogica de Chile (CONICYT). Computations were performed on the GPC supercomputer at the SciNet HPC Consortium. SciNet is funded by the CFI under the auspices of Compute Canada, the Government of Ontario, the Ontario Research Fund -- Research Excellence; and the University of Toronto. Funding from ERC grant 259505 supports JD, EC, and TL. We thank George Efstathiou and Steven Gratton for useful discussions, and Christian Reichardt for help with the SPT data. We acknowledge the use of the Legacy Archive for Microwave Background Data Analysis (LAMBDA). Support for LAMBDA is provided by the NASA Office of Space Science. The likelihood codes will be made public through LAMBDA (\url{http://lambda.gsfc.nasa.gov/}) and the ACT website (\url{http://www.physics.princeton.edu/act/}).

\appendix

\begin{figure*}
\epsscale{1.07}
\plotone{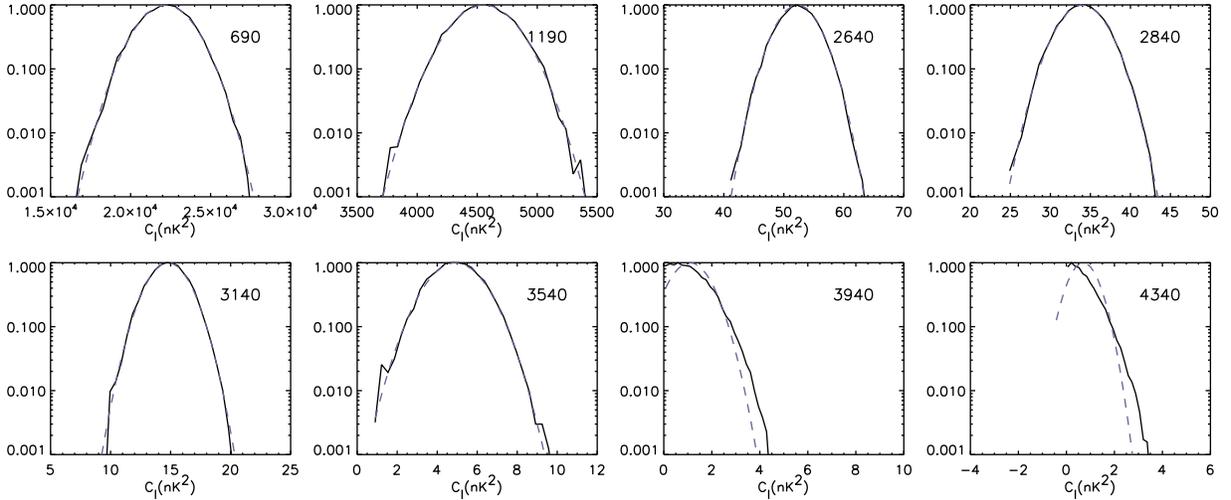}
\caption{ Probability distributions of a selection of CMB bandpowers for ACT-E in the range $600<\ell<4500$, marginalized over secondary parameters (solid). The bin-center for each bandpower shown is indicated on each panel. The bandpowers are well-approximated by Gaussian distributions (dashed), except at scales $\ell \gtsim 3700$. The same behavior is seen for the ACT-S bandpowers.
\label{fig:cl_dist}
}

\end{figure*}

In this Appendix, we perform additional tests on the CMB-only likelihood. In Fig \ref{fig:cl_dist} we show a selection of the distributions of the CMB bandpowers from the estimated  $600<\ell<4500$ range. Distributions for the ACT-E and ACT-S bandpowers are compared to Gaussian distributions (dashed curves); bandpowers at $\ell>3900$ are significantly non-Gaussian, but are well fit by Gaussians at larger scales. The same behaviour is found for the ACT-S bandpowers.

\begin{figure*}
\epsscale{1.05} 
\plotone{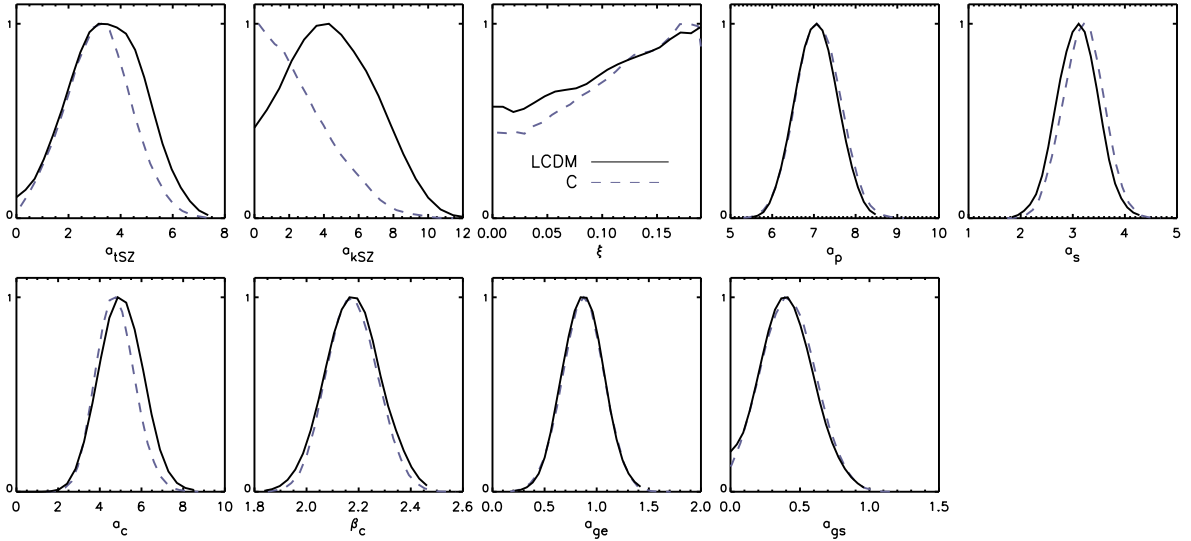}
\caption{Secondary SZ and foreground parameters estimated assuming the \LCDM\ model (`LCDM'), compared to the same parameters estimated jointly with primary CMB bandpowers (`C'). They are consistent, but an anti-correlation between the primary CMB bandpowers and the kSZ power leads to a reduction in $a_{\rm kSZ}$  in the latter case.
\label{fig:params_1d}
}
\end{figure*}

We then compare the secondary parameters estimated in two ways: (1) estimating CMB bandpowers, and (2) estimating 6 \LCDM\  parameters. The distributions are shown in Fig \ref{fig:params_1d}, and are consistent. The tSZ, point source parameters, and Galactic cirrus parameters are not strongly affected by the CMB model assumptions. The kSZ power, $a_{\rm kSZ}$, is $\sim 1\sigma$ lower in the model-independent case, as it is anti-correlated with the CMB bandpowers at $\ell>2000$ due to the common blackbody dependence.  The data cannot distinguish between lensed CMB power and kSZ power at $\ell\sim3000$ scales, so the preference for a smaller kSZ value in the model-indepedent case is driven by the prior that the CMB power is positive.  We test this by allowing the CMB bandpowers to take unphysical negative values. Here, the kSZ power increases to $a_{\rm kSZ}<12$ at 95\% confidence, more consistent with the limits when \LCDM\ is assumed.


\end{document}